\documentclass[11pt,a4paper]{article}
\usepackage{amssymb}
\usepackage{amsmath}
\usepackage{slashed}
\usepackage{bbm}
\usepackage{dsfont}
\usepackage{graphicx}
\usepackage{mathrsfs}
\usepackage{subfig}
\usepackage[english]{babel}
\usepackage{xcolor}
\definecolor{darkblue}{rgb}{0.1,0.1,.7}
\usepackage[breaklinks,colorlinks, linkcolor=darkblue]{hyperref}
\usepackage{tikz}
\usetikzlibrary{intersections}

\newcommand{\be}{\begin{equation}}
\newcommand{\ee}{\end{equation}}
\newcommand{\nn}{\nonumber}

\newcommand{\cD}{{\cal D}}

\newcommand{\cN}{{\cal N}}

\newcommand{\half}{{\tfrac{1}{2}}}

\newcommand{\pr}{\partial}

\newcommand{\bphi}{{\bar \phi}}

\newcommand \bgam{{\bar \gamma}}

\newcommand\bY{{\bar Y}}
\newcommand{\tr}{{\rm tr}}

\newcommand \tbet{\tilde \beta}

\csname @addtoreset\endcsname{equation}{section}
\makeatletter
\def\blfootnote{\xdef\@thefnmark{}\@footnotetext}
\makeatother

\begin{document}
\numberwithin{equation}{section}

\begin{titlepage}
\thispagestyle{empty}
\begin{flushright}
\small
DAMTP-2016-43\\
% ?\\
\today \\
\normalsize
\end{flushright}
\vspace{1cm} 
\begin{center}
{\LARGE {\bf Scheme Dependence and Multiple Couplings}}\\
\end{center}
\vspace{1cm}
\begin{center}
{\bf \Large Ian Jack$^a$ and Hugh Osborn$^b$}%\blfootnote{email:ho@damtp.cam.ac.uk}

\vskip 2cm
${}^a$Department of Mathematical Sciences, University of Liverpool,\\
 {Liverpool L69 3BX}
\vskip 6pt

${}^b$Department of Applied Mathematics and Theoretical Physics, Wilberforce Road, Cambridge CB3 0WA
\end{center}

\date{latest update: \today}
%\maketitle
\begin{abstract}
For theories with multiple couplings the perturbative  $\beta$-functions for scalar, Yukawa couplings
are expressible in terms of contributions corresponding to one particle irreducible graphs and also
contributions which are one particle reducible depending on the anomalous dimension.
Here we discuss redefinitions, or changes of scheme, which preserve this structure.
The redefinitions allow for IPR contributions of a specific form, as is necessary to encompass the
relation between MS and momentum subtraction renormalisation schemes. Many multiply 1PR
terms in the transformed  $\beta$-function are generated but these can all be absorbed into 
antisymmetric contributions to the anomalous dimensions which are essentially arbitrary and 
can be discarded. As an illustration the results are applied to the scheme dependence of the
anomalous dimension, which determines the $\beta$-function, for $\cN=1$ supersymmetric scalar
fermion theories in four dimensions up to four loops.

 \end{abstract}

\thispagestyle{empty}
\end{titlepage}
\pagenumbering{roman}
%\newpage
%\tableofcontents
\newpage

\pagenumbering{arabic}

\setcounter{footnote}{0}

\section{Introduction}

No physical quantities depend on the choice of regularisation scheme.
As is of course well known different schemes are related by reparameterisations
of the finite couplings specifying the renormalised theory. The 
$\beta$-functions, determining the RG flow of the couplings under changes 
of scale, then transform as vector fields under such reparameterisations.

However in perturbative calculations the form of the $\beta$-functions 
are not entirely arbitrary vector fields. 
The counterterms which are added to the original Lagrangian to
ensure finiteness order by order are determined entirely by considerations 
of one particle irreducible, 1PI, graphs whose superficial degree of divergence
is zero or higher. The amplitudes corresponding to one particle reducible, 1PR,
graphs are then automatically finite. Although the local divergent terms
which are subtracted by the counterterms are 1PI the rescaling of the
fields necessary to ensure a canonical normalisation of the kinetic term,
and which is essential in the derivation of RG equations, ensure that 
the $\beta$-functions for scalar or Yukawa couplings have 1PR pieces. 
These 1PR contributions have a very specific form, being determined 
by the anomalous dimension. 

In this paper we address the question of what redefinitions of the couplings
are allowed which preserve the perturbative structure of the $\beta$-functions.
Manifestly for a single coupling there is no way to distinguish 1PI and
1PR contributions to the $\beta$-function. Nevertheless for arbitrary 
couplings each term in the
perturbative $\beta$-function can be identified with particular 1PI and 1PR
graphs. The particular form of the 1PR graphs restricts the resulting 
form for the perturbative $\beta$-functions.

These issues recently came to the fore in a discussion of renormaliseable 
six dimensional $\phi^3$ theory, with a general interaction 
$\frac16\, g_{ijk} \phi_i \phi_j \phi_k$, \cite{Gracey}. In order to relate 
results
for the $\beta$-function when calculated in MS and momentum subtraction schemes
it was crucial to consider 1PR terms in the necessary reparameterisation of 
the couplings. For the perturbative structure of the $\beta$-functions
to be preserved it was necessary to restrict the 1PR contributions to
the redefinition of the couplings to a particular form and to allow
potential antisymmetric 1PR contributions to the anomalous dimension. Here
we discuss such issues in general  and show how the antisymmetric pieces,
which contain 1PR terms, can be dropped and the resulting anomalous
dimension is expressible solely in terms of 1PI graphs with vertices
determined by the transformed coupling.

As an example we consider $\cN=1$ supersymmetric scalar fermion
theories where due to supersymmetric non renormalisation  theorems
the $\beta$-function is determined by the anomalous dimension. This
structure is not preserved by general redefinitions of the couplings but
with the results here we are able to identify scheme changes which
preserve the supersymmetric form. We are then able to determine the
scheme independent contributions at three and four loops. In some cases
these may be determined in terms of lower order results using the $a$-theorem.

In the following section the requirements for changes of  scheme which preserve
 the structure of the $\beta$-function are discussed for both real and complex couplings
  and then these results are supplied to the supersymmetric case in section 3.

\section{Changes of Scheme}

For theories with multi-component fields $\phi_i$ 
and couplings $g^I $
the $\beta$-functions determined by perturbative expansions have
the generic form, with $\gamma_i{}^j(g)$ the anomalous dimension matrix,
\be
\beta^I(g) = \tbet^I(g) + \big (g\,  \gamma(g)  \big ) {}^I \, ,
\label{formB}
\ee
where $\tbet^I$ and $ \gamma_i{}^j$ are constructed from 1PI graphs.

Initially we focus on real fields $\phi_i$ and associated real couplings 
$g^I=\{ g_{i_1 \dots i_n} \}$, taking then $\gamma_i{}^j \to \gamma_{ij}$. 
In this case in \eqref{formB}
\be 
\big (g\,  \gamma(g) \big ) {}_{i_1 \dots i_n} =
g_{j i_2 \dots i_n} \gamma_{j i_1}(g) + \dots +
g_{i \dots i_{n-1} j } \gamma_{ ji_n}(g) \, .
\ee
Acting on the effective action RG flow is generated by
\be
\cD = \cD_\beta
- \int (\gamma(g) \phi)_i \frac{\delta}{\delta \phi_i} \, ,
\label{Drg}
\ee
for
\be
\cD_\beta = \beta^I(g) \frac{\pr}{\pr g^I} = 
\big ( \tbet(g) + \big ( g\, \gamma(g) \big )\big ){}^I 
\frac{\pr}{\pr g^I} \, .
\ee
Crucially there is a arbitrariness under
\be
\gamma(g) \to \gamma(g) + \omega \, , \qquad \omega^T = - \omega \, ,
\ee
since the effective action is invariant under the action of $\cD$ for 
arbitrary choices of $\omega$. If $\omega$ is the generator of a symmetry 
of the theory then the couplings are constrained by $(g \, \omega)^I= 0$.
In consequence, in the context of the RG flow generated by $\cD$,  
$\gamma(g)$ may always be chosen to be symmetric.

We first consider an infinitesimal change in the couplings $\delta g^I$ 
for which
\be
\delta \beta^I  = \beta^J \frac{\pr}{\pr g^J} \, \delta g^I
- \delta g^J \frac{\pr}{\pr g^J} \, \beta^I \, ,
\label{scheme}
\ee 
and assume changes such that
\be 
\delta g^I (g) = f^I(g) + \big ( g\, c(g) \big ) {}^I \, ,
\ee
where $f^I(g), \, c_{ij}(g)$ correspond to one particle irreducible
(1PI) graphs with $n,\, 2$
external lines. 
Initially we do not impose any requirements for $c, \gamma$ to be symmetric.
The variation \eqref{scheme} gives
\be
\delta \beta^I (g) = \delta \tbet^I(g) + 
\big ( g\, \delta \gamma(g) \big ) {}^I + 
\big ( g\, \omega(g)  \big ) {}^I \, , \quad
\omega(g) = \gamma(g)^T c(g) - c(g)^T \gamma(g)   \, ,
\label{dbeta}
\ee
with
\begin{align}
 \delta \tbet^I(g) 
= {}& \big ( \tbet(g) + 
\big ( g\, \gamma(g)  \big )\big ){}^J 
\frac{\pr}{\pr g^J}  \, f^I(g)
- \big ( f(g) + \big ( g \, c(g) \big ) \big ) {}^J 
\frac{\pr}{\pr g^J} \, \tbet^I(g) \nn \\
&{} - \big (f(g) \, \gamma(g) \big ){}^I +  
\big ( \tbet(g)  \, c(g) \big ) {}^I \, ,
\label{deltaB}
\end{align}
and
\begin{align}
\delta \gamma(g)  = {}& \big ( \tbet(g) + 
\big ( g\, \gamma(g) \big )\big ){}^J 
\frac{\pr}{\pr g^J} \, c(g) - 
\big ( f(g) + \big ( g \, c(g) \big ) \big ){}^J \,  
\frac{\pr}{\pr g^J} \, \gamma(g) \nn \\
&{}+ \big [ \gamma(g) , \, c(g) \big ] - \omega(g) \, .
\label{deltag}
\end{align}
Crucially, with the assumptions on the form of $f^I, \, c$,
$\delta \tbet^I , \, \delta \gamma$ correspond just to the 
contributions of 1PI graphs (the second lines of \eqref{deltaB}, \eqref{deltag} remove
1PR terms arising from the differentiation in the first line acting
on the couplings which couple to the external lines).
In particular $\big ( \gamma(g) \, g \big ){}^J \pr_J \, c(g)  - 
c(g) \, \gamma(g) - \gamma(g)^T c(g)$ is 1PI so long as  $c$ is 1PI. A similar
result holds for $c\leftrightarrow \gamma$. 
If $c,\gamma$ are symmetric then so is $\delta \gamma$ and the last line of
\eqref{deltag} vanishes.

The extension of \eqref{dbeta} to finite redefinitions of the couplings, 
or general scheme changes, requires the $\beta$-function to transform as
a tangent vector to the manifold parameterised by the couplings and hence
\be
\beta'{}^I(g') = \beta^J(g) \frac{\pr}{\pr g^J} \, g'{}^I(g) \quad
\mbox{for} \quad g^I \to  g'{}^I(g) \, .
\label{varB}
\ee
In perturbation theory $g'{}^I$ is required to have  a power series expansion 
in $g$ and to lowest order $g'{}^I(g) = g^I + {\rm O}(g^2)$. 
For redefinitions which preserve the structure \eqref{formB}, it is necessary that 
the transformed $\beta$-function has the form
\be
\beta'{}^I(g') = \tbet'{}^I(g') + 
\big ( g'\, (\gamma'(g') + \Omega) \big ) {}^I  \, ,
\qquad \Omega^T = - \Omega \, ,
\label{vv}
\ee
with $\tbet'{}^I(g') , \, \gamma'(g')$ expressible in terms of 1PI
contributions while $\Omega$ is expressed in terms of 1PR graphs.
We define first $(g\circ C)^I$ where
\be
( g \circ C)_{i_1 \dots i_n} = g_{j_1 \dots j_n} \, C_{j_1 i_1} \dots
C_{j_n i_n} \, .
\ee
If $C= \mathds{1} + c + \dots $ then 
$(g\circ C)^I = g^I + (g \, c )^I + \dots$. We then
consider changes of scheme of the form
\be
g'{}^I = \big (  (g+f(g) ) \circ C(g)  \big ){}^I \, ,
\label{gp}
\ee
where $f^I(g)$ is expressed in terms of 1PI graphs with $n$ external lines.
$C$ is also determined in terms of 1PI graphs later, $( g+f) \circ C $
generates 1PR terms in the redefinition of the couplings. Using \eqref{gp} in
\eqref{varB} we obtain
\begin{align}
\tbet'{\hskip 0.5pt}^I(g') = {}& \big (  \tbet_f(g+f) \circ C \big ){}^I \, ,
\nn \\
\tbet_f{}^I (g+f)  = {}& \tbet^I(g) + \cD_\beta f^I(g)
- \big ( f (g) \, \gamma(g) \big ){}^I \, .
\label{finB} 
\end{align}
$\tbet_f{}^I (g) $ is expressible as a sum of 1PI contributions as a function
of $g$, $\tbet_f(g+f)\circ C $ ensures that $\tbet'(g')$ is also given
in terms of 1PI graphs as a function of $g'$.  For $C= \mathds{1} + c$ and
$c,f$ infinitesimal \eqref{finB} reduces to \eqref{deltaB} for
$\delta \tbet^I(g) = \tbet'{\hskip 0.5pt}^I(g) - \tbet^I(g)$. Furthermore
\eqref{vv} is then satisfied if we take
\be
\gamma'(g') + \Omega
= C^{-1} \big ( \cD_\beta + \gamma(g) \big ) C \, .
\label{ngam}
\ee
As argued above any antisymmetric $\Omega$ is irrelevant when considering
the action of $\cD$ in \eqref{Drg}.

To analyse this further we assume $\gamma^T = \gamma$ and 
\begin{align}
C(g) = {}& \mathds{1} + c(g) + a_2 \, c(g)^2 + a_3 \, c(g)^3  + \dots \, , \quad c^T = c \, ,\nn \\
C(g)^{-1} ={}& \mathds{1} - c(g) - ( a_2-1)  \, c(g)^2 - (a_3- 2a_2 +1)  \, c(g)^3  + \dots 
\end{align}
with $c(g)$ expressible in terms of a sum of contributions 
corresponding to 1PI graphs. We also define 
\be
\cD_\beta \, c(g) =
\gamma(g) \, c(g) + c(g) \, \gamma(g) + c_\beta(g) \, ,
\label{diffc}
\ee
where $c_\beta(g)$ is expressible in terms of 1PI graphs if $c(g)$ is 1PI.
From \eqref{ngam} we then have
\begin{align}
C(g)^{-1} \big (  & \gamma'(g') + \Omega \big ) C(g)^{-1}  \nn \\
= {}&
\gamma(g) + c_\beta(g) + \gamma(g)\, c(g) - c(g)\,  \gamma(g) \nn \\
&{}+  (a_2-1) \big ( \gamma(g) \, c(g)^2 - c(g)^2 \gamma(g) \big ) 
+ (2a_2 -3) \, c(g) \gamma(g) c(g) \nn \\
&{}+ (a_2-1) \, c_\beta(g)c(g) + (a_2-2) \, c(g) c_\beta(g) \nn \\
&{}+ (a_3 -2a_2 +1) 
\big ( \gamma(g) \, c(g)^3 + c_\beta(g) c(g)^2 - c(g)^3 \gamma(g) \big ) \nn \\
&{}+ (2a_3-5a_2 +3) \, c(g) \gamma(g) c(g)^2 +
 (2a_3- 7a_2 +5) \, c(g)^2 \gamma(g) c(g) \nn \\
&{}+ (a_3 - 3a_2 +2)\, c(g) c_\beta(g) c(g) + (a_3 -4a_2 +3) \, c(g)^2 c_\beta(g)
+ {\rm O}(c^4) \, .
\end{align}
Choosing
\be
a_2 = \tfrac32 \, , \qquad a_3 = \tfrac52 \, ,
\ee
gives
\begin{align}
C(g)^{-1} \big (  & \gamma'(g') + \Omega \big ) C(g)^{-1}  \nn \\
= {}&
\gamma(g) + c_\beta(g) + \gamma(g)\, c(g) - c(g)\,  \gamma(g) \nn \\
&{}+  \half \big ( \gamma(g) \, c(g)^2 - c(g)^2 \gamma(g) +
c_\beta(g)c(g) -  c(g) c_\beta(g) \big )  \nn \\
&{}+ \tfrac12 
\big ( \gamma(g) \, c(g)^3 - c(g)^3 \gamma(g) +
c(g) \gamma(g) c(g)^2 - c(g)^2 \gamma(g) \big ) \nn \\
&{}+ \big ( c_\beta(g) c(g)^2 - c(g)^2 c_\beta(g) \big ) 
+ {\rm O}(c^4) \, .
\label{res4}
\end{align}
This allows us to identify
\be
\gamma'(g') = C(g) \big ( \gamma(g) + c_\beta(g) \big ) C(g) \, ,
\label{gamp}
\ee
where, since $\gamma(g),c_\beta(g)$ correspond to 1PI graphs as functions of
$g$, then $\gamma'(g')$ is also given in terms of 1PI graphs as a function
of $g'$. The remaining part of \eqref{res4} determines $\Omega=-\Omega^T$
as a sum of 1PR contributions.

To extend these results beyond the first few terms in the expansion of $C$ 
it is sufficient to take
\begin{align}
C(g) = {}& \big(\mathds{1} - 2\, c(g) \big )^{-\frac12} 
= \sum_{n=0}^\infty \frac{1}{n!} \, (\half)_n \, (2c)^n \, ,\nn \\
C(g)^{-1} = {}& \big(\mathds{1} - 2\, c(g) \big )^{\frac12} 
= \sum_{n=0}^\infty \frac{1}{n!} \, (-\half)_n \, (2c)^n \, .
\label{CCdef}
\end{align}
Then inserting \eqref{gamp} in \eqref{ngam}, using \eqref{diffc} 
and $C(g)^{-2} = 1 - 2\, c(g)$, gives
\begin{align}
C(g)^{-1} \Omega \, C(g)^{-1} = {}& 
\gamma(g)\, c(g) - c(g)\,  \gamma(g) \nn \\
&{} + \tfrac12 \, \cD_\beta  C(g)^{-1}  \; C(g)^{-1} -
\tfrac12 \, C(g)^{-1} \, \cD_\beta C(g)^{-1}  \, , 
\label{general}
\end{align}
from which it is evident that, with this choice of $C(g)$,
$\Omega^T = - \Omega$. Expanding $\cD_\beta C^{-1}$ gives
in \eqref{general}
\begin{align}
C(g)^{-1} \Omega \, C(g)^{-1} = {}& 
\gamma(g)\, c(g) - c(g)\,  \gamma(g) \nn \\
&{} + \sum_{N=2}^\infty \bigg ( \sum_{r=0}^N \, 
f_{N,r} \, c(g)^r  \gamma(g) \, c(g)^{N-r} + 
 \sum_{r=0}^{N-1}  g_{N,r} \, 
 c(g)^r  c_\beta(g) \, c(g)^{N-1-r} \bigg ) \, , 
\label{general2}
\end{align}
where, for $N\ge 2$,
\be
g_{N,r} = \sum_{m=0}^{N-1-r} h_{N,m} \, , \quad
f_{N,r} = 2 \, g_{N,r-1} - h_{N,r} \, , \qquad
h_{N,r} = 2^N \frac{(-\tfrac12)_r \, (-\tfrac12)_{N-r}} {r! \, (N-r)!} \, .
\label{gdef}
\ee
and we may take $g_{N,-1}=0$.

If we define
\be
{c}'(g') = - C(g) \, c(g) \, C(g) \, ,
\label{cdash}
\ee
then ${c}'(g')$ is defined in terms of 1PI graphs so long as $c(g)$ is,
and from \eqref{CCdef}
\be
{C}'(g') = C(g)^{-1} = 
\big(\mathds{1} - 2\, {c}'(g') \big )^{-\frac12} \, .
\ee
The inverse of \eqref{gp} is then
\be
g{}^I = \big ( ( g' + {f}'(g')) \circ { C}'(g') \big ){}^I \, ,
\quad {f}'(g')  = - f(g) \circ C(g) \, ,
\label{gpi}
\ee
with $ {f}'(g') $ determined by 1PI graphs.

These results may easily be extended to complex fields $\phi_i, \, \bphi^i$
and complex couplings $g^I = \{ g_{i_1 \dots i_p}{}^{j_1 \dots j_q} \}$. 
In this case
\be
\beta^I(g) = \tbet^I(g) + \big (g\,  \gamma(g)  \big ) {}^I +
\big ( \bgam(g) \, g \big ){}^I \, ,
\label{formC}
\ee
for 
\begin{align}
\big (g\,  \gamma(g) \big ){}_{i_1 \dots i_p}{}^{j_1 \dots j_q}  = {}&
g_{i_i \dots i_p}{\,}^{k \, j_2 \dots j_q} \, \gamma_{k}{}^{j_1}(g) + \dots +
g_{i \dots i_{p} }{\,}^{j_1 \dots j_{q-1}\, j_k } \, 
\gamma_k{}^{j_q}(g)\, \nn \\
\big (\bgam(g) \, g \big ){}_{i_1 \dots i_p}{}^{j_1 \dots j_q} = {}&
\bgam_{i_1}{}^k (g) \, g_{k\, i_2 \dots i_p}{\,}^{j_1 \dots j_q} + \dots +
\bgam_{i_p}{}^k (g) \, g_{i_1 \dots i_{p-1} \, k }{\,}^{j_1 \dots j_q}  \, ,
\label{bgamc}
\end{align}
and $\bgam = \gamma^\dagger$.
In this case \eqref{Drg} becomes
\be
\cD = \cD_\beta
- \int (\gamma(g) \phi)_i \frac{\delta}{\delta \phi_i} 
- \int (\bphi \hskip 0.8pt \bgam(g) )^i \frac{\delta}{\delta \bphi^i} \, ,
\label{Drgc}
\ee
where $\gamma \to \gamma + \omega , \, \bgam \to \bgam - \omega$ corresponds
to a trivial invariance of the effective action. Since the $\omega$ terms 
can be discarded we may impose $\bgam = \gamma$. To preserve the 
perturbative structure in \eqref{bgamc} we may extend \eqref{gp} to
\be
g'{}^I = \big (  C(g) \circ (g+f(g) ) \circ C(g)  \big ){}^I \, ,
\label{gpc}
\ee
with $f(g)$ representing 1PI contributions to the redefinition of
the couplings and $C_i{}^j(g)$ expressible as in \eqref{CCdef} in terms of
the 1PI $c(g) = {\bar c}(g) = c(g)^\dagger$. In this case \eqref{finB} becomes
\begin{align}
\tbet'{\hskip 0.5pt}^I(g') = {}& 
\big (  C(g) \circ \tbet_f(g+f) \circ C(g) \big ){}^I \, ,
\nn \\
\tbet_f{}^I (g+f)  = {}& \tbet^I(g) + \cD_\beta f^I(g)
- \big ( f (g) \, \gamma(g) \big ){}^I \, -
\big ( \gamma(g)\, f (g) \big ){}^I \ ,
\label{finBic} 
\end{align}
and \eqref{ngam} becomes
\be
\gamma'(g') + \Omega = C(g)^{-1} \big ( \cD_\beta + \gamma(g) \big ) 
C(g) \, ,\quad
\gamma'(g') - \Omega =  \big ( \cD_\beta C(g) + C(g) \gamma(g) \big ) \,  
C(g)^{-1} \, , 
\label{ngamc}
\ee
with $\Omega = - \Omega^\dagger$. Otherwise the discussion follows the 
same route as before. By acting on the left and the right of \eqref{ngamc}
with $C^{-1}$ and subtracting the two equations we recover \eqref{general}.
The modified $\gamma'(g')$ is determined by \eqref{gamp} and \eqref{diffc}
as before.

\section{Supersymmetric Model}

A cardinal illustration of the relevance of  these issues is the $\cN=1$
supersymmetric scalar fermion theory in four dimensions. The complex couplings
are then symmetric rank three tensors $Y^{ijk}, \, \bY_{ijk}$ and there
are no 1PI contributions to the $\beta$-functions which are determined just by
the anomalous dimension $\gamma_i{}^j(Y,\bY)$ in the form
\be
\beta_Y{\hskip -0.5pt}^{ijk} = ( Y \gamma )^{ijk} \, , \qquad
\beta_{\bY}{}_{ijk} = ( \gamma \, \bY)_{ijk} \, ,
\label{Ybet}
\ee
with notation as in \eqref{bgamc}. Scheme changes at the linearised level
preserving \eqref{Ybet}  
for this theory were discussed in \cite{Jack} (see 7.23) and these issues 
were also considered in \cite{Jack3}. More generally from the
results above the form \eqref{Ybet} is consistent with  redefinitions
\be
Y'{}^{ijk} = Y^{lmn} \, C_l{}^i C_m{}^j C_n{}^k \, , \qquad
\bY'{\!}_{ijk} = C_i{}^l C_j{}^m C_k{}^n \, \bY_{lmn} \, ,
\label{YCCC}
\ee
where $C_i{}^l(Y,\bY)$ is determined in terms of a 1PI $c_i{}^l(Y,\bY)$
as in \eqref{CCdef}.

Applying these results first to three loops we let $Y^{ijk} \to 4\pi \, Y^{ijk} , \,
\bY_{ijk} \to 4\pi \, \bY_{ijk}$ and then use a diagrammatic 
notation where 
$Y^{ijk} \to 
\tikz[baseline=(vert_cent.base)]{
  \node (vert_cent) {\hspace{-13pt}$\phantom{-}$};
  \draw (0,0.3)--(0.3,0)
        (0,-0.3)--(0.3,0)
        (0.3,0)--(0.65,0);
}
$ and
$\bY_{ijk} \to
\tikz[baseline=(vert_cent.base)]{
  \node (vert_cent) {\hspace{-13pt}$\phantom{-}$};
  \draw (-0.1,0)--(0.3,0)
        (0.3,0)--(0.6,0.3)
        (0.3,0)--(0.6,-0.3);
          \filldraw [gray] (0.3,0) circle [radius=1.5pt];
}
$.  Only lines linking $Y$ and $\bY$ vertices are allowed. At this order
\begin{align}
\gamma(Y,\bY) &{} = c_1 \, 
\tikz[baseline=(vert_cent.base)]{
  \node (vert_cent) {\hspace{-13pt}$\phantom{-}$};
  \draw (-0.4,0)--(0.1,0)
        (0.7,0) ++(0:0.6cm) arc (0:360:0.6cm and 0.4cm)
        (1.3,0)--(1.8,0);
        \filldraw [gray] (0.1,0) circle [radius=1.5pt];
}
+ c_2\, 
\tikz[baseline=(vert_cent.base)]{
  \node (vert_cent) {\hspace{-13pt}$\phantom{-}$};
  \draw (-0.4,0)--(0.1,0)
       (0.7,0) ++(0:0.6cm and 0.4cm) arc (0:50:0.6cm and 0.4cm) node (n1)
        {}
       (0.7,0) ++(50:0.6cm and 0.4cm) arc (50:130:0.6cm and 0.4cm) node (n2) 
       {}
       (0.7,0) ++(130:0.6cm and 0.4cm) arc (130:360:0.6cm and 0.4cm)
        (n1.base) to[out=215,in=325] (n2.base)
                (1.3,0)--(1.8,0);
                  \filldraw [gray] (n1) circle [radius=1.5pt]; 
          \filldraw [gray]  (0.1,0) circle [radius=1.5pt]; 
          }
\nn \\ 
&{} + c_{3A} \,
\tikz[baseline=(vert_cent.base)]{
  \node (vert_cent) {\hspace{-13pt}$\phantom{-}$};
  \draw (-0.4,0)--(0.1,0)
     (0.7,0) ++(0:0.6cm and 0.4cm) arc (0:35:0.6cm and 0.4cm) node (n1)
            {}
     (0.7,0) ++(35:0.6cm and 0.4cm) arc (35:75:0.6cm and 0.4cm) node(n2)
       {}
        (0.7,0) ++(75:0.6cm and 0.4cm) arc (75:105:0.6cm and 0.4cm) node(n3){}
      (n1.base) to  [out=270,in=240] (n2.base) 
          (0.7,0) ++(105:0.6cm and 0.4cm) arc (105:145:0.6cm and 0.4cm)  node(n4) {} 
          (0.7,0) ++(145:0.6cm and 0.4cm) arc (145:360:0.6cm and 0.4cm) 
       (n3.base) to  [out= 330,in= 300] (n4.base) 
         (1.3,0)--(1.8,0);
                   \filldraw [gray] (n1) circle [radius=1.5pt]; 
          \filldraw [gray]  (0.1,0) circle [radius=1.5pt];
          \filldraw [gray] (n3) circle [radius=1.5pt]; 
          }
+ c_{3B} \,
\tikz[baseline=(vert_cent.base)]{[scale=1.5]
  \node (vert_cent) {\hspace{-13pt}$\phantom{-}$};
  \draw (-0.4,0)--(0.1,0)
     (0.7,0) ++(0:0.6cm and 0.4cm) arc (0:50:0.6cm and 0.4cm) node (n1)
            {}
     (0.7,0) ++(50:0.6cm and 0.4cm) arc (50:130:0.6cm and 0.4cm) node(n2)
       {}
        (0.7,0) ++(130:0.6cm and 0.4cm) arc (130:230:0.6cm and 0.4cm) node(n3){}
          (n1.base) to[out=215,in=325] (n2.base) 
          (0.7,0) ++(230:0.6cm and 0.4cm) arc (230:310:0.6cm and 0.4cm)  node(n4) {} 
          (0.7,0) ++(310:0.6cm and 0.4cm) arc (310:360:0.6cm and 0.4cm) 
         (n3.base) to[out=-325,in=-215] (n4.base) 
         (1.3,0)--(1.8,0);
                   \filldraw [gray] (n1) circle [radius=1.5pt]; 
          \filldraw [gray]  (0.1,0) circle [radius=1.5pt];
          \filldraw [gray] (n4) circle [radius=1.5pt]; 
          } 
%\nn \\ &
{}+ c_{3C}  \,
\tikz[baseline=(vert_cent.base)]{
  \node (vert_cent) {\hspace{-13pt}$\phantom{-}$};
  \draw (-0.4,0)--(0.1,0)
     (0.7,0) ++(0:0.6cm and 0.4cm) arc (0:35:0.6cm and 0.4cm) node (n1)
            {}
     (0.7,0) ++(35:0.6cm and 0.4cm) arc (35:65:0.6cm and 0.4cm) node(n2)
       {}
        (0.7,0) ++(65:0.6cm and 0.4cm) arc (65:115:0.6cm and 0.4cm) node(n3){}
          (0.7,0) ++(115:0.6cm and 0.4cm) arc (115:145:0.6cm and 0.4cm)  node(n4) {} 
          (0.7,0) ++(145:0.6cm and 0.4cm) arc (145:360:0.6cm and 0.4cm) 
       (n2.base) to  [out= 220,in= 320] (n3.base) 
       (n1.base) to  [out=220,in=320] (n4.base) 
         (1.3,0)--(1.8,0);
                   \filldraw [gray] (n1) circle [radius=1.5pt]; 
          \filldraw [gray]  (0.1,0) circle [radius=1.5pt];
          \filldraw [gray] (n3) circle [radius=1.5pt]; 
          }
+ c_{3D} \,
\tikz[baseline=(vert_cent.base)]{
  \node (vert_cent) {\hspace{-13pt}$\phantom{-}$};
  \draw (-0.4,0)--(0.1,0)
     (0.7,0) ++(0:0.6cm and 0.4cm) arc (0:50:0.6cm and 0.4cm) node (n1)
            {}
     (0.7,0) ++(50:0.6cm and 0.4cm) arc (50:130:0.6cm and 0.4cm) node(n2)
       {}
        (0.7,0) ++(130:0.6cm and 0.4cm) arc (130:230:0.6cm and 0.4cm) node(n3){}
          (0.7,0) ++(230:0.6cm and 0.4cm) arc (230:310:0.6cm and 0.4cm)  node(n4) {} 
          (0.7,0) ++(310:0.6cm and 0.4cm) arc (310:360:0.6cm and 0.4cm) 
           (n1.base) to (0.75,0.05)
         (n2.base) to  (n4.base) 
         (0.65,-0.05) to (n3.base)
         (1.3,0)--(1.8,0);
                   \filldraw [gray] (n1) circle [radius=1.5pt]; 
          \filldraw [gray]  (0.1,0) circle [radius=1.5pt];
          \filldraw [gray] (n4) circle [radius=1.5pt]; 
          }
          .
          \label{three}
\end{align}
We consider to this order transformations determined by
 \be
c(Y,\bY) = \epsilon_1 \, 
\tikz[baseline=(vert_cent.base)]{
  \node (vert_cent) {\hspace{-13pt}$\phantom{-}$};
  \draw (-0.4,0)--(0.1,0)
        (0.7,0) ++(0:0.6cm) arc (0:360:0.6cm and 0.4cm)
        (1.3,0)--(1.8,0);
        \filldraw [gray] (0.1,0) circle [radius=1.5pt];
}
+  \epsilon_2 \, 
\tikz[baseline=(vert_cent.base)]{
  \node (vert_cent) {\hspace{-13pt}$\phantom{-}$};
  \draw (-0.4,0)--(0.1,0)
       (0.7,0) ++(0:0.6cm and 0.4cm) arc (0:50:0.6cm and 0.4cm) node (n1)
        {}
       (0.7,0) ++(50:0.6cm and 0.4cm) arc (50:130:0.6cm and 0.4cm) node (n2) 
       {}
       (0.7,0) ++(130:0.6cm and 0.4cm) arc (130:360:0.6cm and 0.4cm)
        (n1.base) to[out=215,in=325] (n2.base)
                (1.3,0)--(1.8,0);
                  \filldraw [gray] (n1) circle [radius=1.5pt]; 
          \filldraw [gray]  (0.1,0) circle [radius=1.5pt]; 
          } \, ,
          \label{ctwo}
\ee
where correspondingly from \eqref{cdash}
 \be
c'(Y',\bY') = - \epsilon_1 \, 
\tikz[baseline=(vert_cent.base)]{
  \node (vert_cent) {\hspace{-13pt}$\phantom{-}$};
  \draw (-0.4,0)--(0.1,0)
        (0.7,0) ++(0:0.6cm) arc (0:360:0.6cm and 0.4cm)
        (1.3,0)--(1.8,0);
        \filldraw [gray] (0.1,0) circle [radius=1.5pt];
}
-  (\epsilon_2 - 4 \, \epsilon_1{\!}^2 ) \, 
\tikz[baseline=(vert_cent.base)]{
  \node (vert_cent) {\hspace{-13pt}$\phantom{-}$};
  \draw (-0.4,0)--(0.1,0)
       (0.7,0) ++(0:0.6cm and 0.4cm) arc (0:50:0.6cm and 0.4cm) node (n1)
        {}
       (0.7,0) ++(50:0.6cm and 0.4cm) arc (50:130:0.6cm and 0.4cm) node (n2) 
       {}
       (0.7,0) ++(130:0.6cm and 0.4cm) arc (130:360:0.6cm and 0.4cm)
        (n1.base) to[out=215,in=325] (n2.base)
                (1.3,0)--(1.8,0);
                  \filldraw [gray] (n1) circle [radius=1.5pt]; 
          \filldraw [gray]  (0.1,0) circle [radius=1.5pt]; 
          } \, .
          \label{cthreeb}
\ee
From the definition \eqref{diffc}
\begin{align}
c_\beta(Y,\bY) &{} = 4\,  c_1 \epsilon_1 \, \tikz[baseline=(vert_cent.base)]{
  \node (vert_cent) {\hspace{-13pt}$\phantom{-}$};
  \draw (-0.4,0)--(0.1,0)
       (0.7,0) ++(0:0.6cm and 0.4cm) arc (0:50:0.6cm and 0.4cm) node (n1)
        {}
       (0.7,0) ++(50:0.6cm and 0.4cm) arc (50:130:0.6cm and 0.4cm) node (n2) 
       {}
       (0.7,0) ++(130:0.6cm and 0.4cm) arc (130:360:0.6cm and 0.4cm)
        (n1.base) to[out=215,in=325] (n2.base)
                (1.3,0)--(1.8,0);
                  \filldraw [gray] (n1) circle [radius=1.5pt]; 
          \filldraw [gray]  (0.1,0) circle [radius=1.5pt]; 
          }
\nn \\ 
&{} + 4\,  c_1 \epsilon_2 \, 
\tikz[baseline=(vert_cent.base)]{
  \node (vert_cent) {\hspace{-13pt}$\phantom{-}$};
  \draw (-0.4,0)--(0.1,0)
     (0.7,0) ++(0:0.6cm and 0.4cm) arc (0:35:0.6cm and 0.4cm) node (n1)
            {}
     (0.7,0) ++(35:0.6cm and 0.4cm) arc (35:75:0.6cm and 0.4cm) node(n2)
       {}
        (0.7,0) ++(75:0.6cm and 0.4cm) arc (75:105:0.6cm and 0.4cm) node(n3){}
      (n1.base) to  [out=260,in=260] (n2.base) 
          (0.7,0) ++(105:0.6cm and 0.4cm) arc (105:145:0.6cm and 0.4cm)  node(n4) {} 
          (0.7,0) ++(145:0.6cm and 0.4cm) arc (145:360:0.6cm and 0.4cm) 
       (n3.base) to  [out= 330,in= 320] (n4.base) 
         (1.3,0)--(1.8,0);
                   \filldraw [gray] (n1) circle [radius=1.5pt]; 
          \filldraw [gray]  (0.1,0) circle [radius=1.5pt];
          \filldraw [gray] (n3) circle [radius=1.5pt]; 
          }
+2\, c_1  \epsilon_2 \, 
\tikz[baseline=(vert_cent.base)]{[scale=1.5]
  \node (vert_cent) {\hspace{-13pt}$\phantom{-}$};
  \draw (-0.4,0)--(0.1,0)
     (0.7,0) ++(0:0.6cm and 0.4cm) arc (0:50:0.6cm and 0.4cm) node (n1)
            {}
     (0.7,0) ++(50:0.6cm and 0.4cm) arc (50:130:0.6cm and 0.4cm) node(n2)
       {}
        (0.7,0) ++(130:0.6cm and 0.4cm) arc (130:230:0.6cm and 0.4cm) node(n3){}
          (n1.base) to[out=215,in=325] (n2.base) 
          (0.7,0) ++(230:0.6cm and 0.4cm) arc (230:310:0.6cm and 0.4cm)  node(n4) {} 
          (0.7,0) ++(310:0.6cm and 0.4cm) arc (310:360:0.6cm and 0.4cm) 
         (n3.base) to[out=-325,in=-215] (n4.base) 
         (1.3,0)--(1.8,0);
                   \filldraw [gray] (n1) circle [radius=1.5pt]; 
          \filldraw [gray]  (0.1,0) circle [radius=1.5pt];
          \filldraw [gray] (n4) circle [radius=1.5pt]; 
          } 
%\nn \\ &
{}+ 4(c_1 \epsilon_2  + c_2 \epsilon_1)  \, 
\tikz[baseline=(vert_cent.base)]{
  \node (vert_cent) {\hspace{-13pt}$\phantom{-}$};
  \draw (-0.4,0)--(0.1,0)
     (0.7,0) ++(0:0.6cm and 0.4cm) arc (0:35:0.6cm and 0.4cm) node (n1)
            {}
     (0.7,0) ++(35:0.6cm and 0.4cm) arc (35:65:0.6cm and 0.4cm) node(n2)
       {}
        (0.7,0) ++(65:0.6cm and 0.4cm) arc (65:115:0.6cm and 0.4cm) node(n3){}
          (0.7,0) ++(115:0.6cm and 0.4cm) arc (115:145:0.6cm and 0.4cm)  node(n4) {} 
          (0.7,0) ++(145:0.6cm and 0.4cm) arc (145:360:0.6cm and 0.4cm) 
       (n2.base) to  [out= 220,in= 320] (n3.base) 
       (n1.base) to  [out=220,in=320] (n4.base) 
         (1.3,0)--(1.8,0);
                   \filldraw [gray] (n1) circle [radius=1.5pt]; 
          \filldraw [gray]  (0.1,0) circle [radius=1.5pt];
          \filldraw [gray] (n3) circle [radius=1.5pt]; 
          }
          \, ,
\end{align}
and then applying the transformation $Y,\bY \to Y',\bY'$ to 
$\gamma(Y,\bY) + c_\beta(Y,\bY)$ gives according to \eqref{gamp} 
$\gamma'(Y',\bY')$ as a sum of 1PI terms of the same form as \eqref{three}
with the one and two loop contributions invariant and at three loops
\be
c' {\!}_{3A} = c_{3A} + 2\, X  \, , \quad c'{\!}_{3B} = c_{3B} +  X  \, , \qquad
X = -4c_1 \, \epsilon_1{\!}^2 + 2\, c_1\, \epsilon_2 - 2 \, c_2 \, \epsilon_1 \, , 
\ee
with $c_{3C}, \, c_{3D}$ also invariant. Calculations \cite{Jack3} give for the
scheme invariants to this order
\be
c_1 = \tfrac12 \, , \quad c_2 = - \tfrac12\, , \quad 
c_{3A} - 2 \, c_{3B} =0 \, , \quad  c_{3C} = 1 \, , \quad c_{3D} = \tfrac32 \, \zeta(3) \, .
\ee

The four loop contributions to the anomalous dimension $\gamma$ have significantly
more terms and it is convenient to separate them into three classes according to
the topology of the corresponding diagrams. For planar graphs
\begin{align}
\gamma(Y,\bY)^{(4)}_1 ={}& c_{4A} \, 
\tikz[baseline=(vert_cent.base)]{
  \node (vert_cent) {\hspace{-13pt}$\phantom{-}$};
  \draw (-0.6,0)--(-0.1,0)
     (0.7,0) ++(0:0.8cm and 0.5cm) arc (0:25:0.8cm and 0.5cm) node (n1)
            {}
     (0.7,0) ++(25:0.8cm and 0.5cm) arc (25:60:0.8cm and 0.5cm) node(n2)
       {}
        (0.7,0) ++(60:0.8cm and 0.5cm) arc (60:72.5:0.8cm and 0.5cm) node(n3){}
      (n1.base) to  [out=270,in=250] (n2.base) 
        (0.7,0) ++(72.5:0.8cm and 0.5cm) arc (72.5:107.5:0.8cm and 0.5cm) node(n4){}
          (0.7,0) ++(107.5:0.8cm and 0.5cm) arc (107.5:120:0.8cm and 0.5cm)  node(n5) {} 
          (0.7,0) ++(120:0.8cm and 0.5cm) arc (120:155:0.8cm and 0.5cm) node(n6) {}
            (0.7,0) ++(155:0.8cm and 0.5cm) arc (155:360:0.8cm and 0.5cm)
             (n3.base) to  [out= 280,in= 260] (n4.base) 
     (n5.base) to  [out= 320,in= 300] (n6.base) 
         (1.5,0)--(2,0);
                   \filldraw [gray] (n1) circle [radius=1.5pt]; 
          \filldraw [gray]  (-0.1,0) circle [radius=1.5pt];
          \filldraw [gray] (n3) circle [radius=1.5pt]; 
           \filldraw [gray] (n5) circle [radius=1.5pt]; 
          }
+ c_{4B} \, 
\tikz[baseline=(vert_cent.base)]{
  \node (vert_cent) {\hspace{-13pt}$\phantom{-}$};
  \draw (-0.6,0)--(-0.1,0)
     (0.7,0) ++(0:0.8cm and 0.5cm) arc (0:35:0.8cm and 0.5cm) node (n1)
            {}
     (0.7,0) ++(35:0.8cm and 0.5cm) arc (35:75:0.8cm and 0.5cm) node(n2)
       {}
        (0.7,0) ++(75:0.8cm and 0.5cm) arc (75:105:0.8cm and 0.5cm) node(n3){}
      (n1.base) to  [out=270,in=240] (n2.base) 
          (0.7,0) ++(105:0.8cm and 0.5cm) arc (105:145:0.8cm and 0.5cm)  node(n4) {} 
          (0.7,0) ++(145:0.8cm and 0.5cm) arc (145:240:0.8cm and 0.5cm) node(n5){}
          (0.7,0) ++(240:0.8cm and 0.5cm) arc (240:300:0.8cm and 0.5cm) node(n6){}
          (0.7,0) ++(300:0.8cm and 0.5cm) arc (300:360:0.8cm and 0.5cm) 
       (n3.base) to  [out= 280,in= 320] (n4.base) 
       (n5.base) to  [out= 80,in= 100] (n6.base) 
         (1.5,0)--(2,0);
                   \filldraw [gray] (n1) circle [radius=1.5pt]; 
          \filldraw [gray]  (-0.1,0) circle [radius=1.5pt];
          \filldraw [gray] (n3) circle [radius=1.5pt]; 
           \filldraw [gray] (n6) circle [radius=1.5pt];
           }
           \nn \\
           &{}+ c_{4C} \left (
           \tikz[baseline=(vert_cent.base)]{
  \node (vert_cent) {\hspace{-13pt}$\phantom{-}$};
  \draw (-0.6,0)--(-0.1,0)
     (0.7,0) ++(0:0.8cm and 0.5cm) arc (0:25:0.8cm and 0.5cm) node (n1)
            {}
     (0.7,0) ++(25:0.8cm and 0.5cm) arc (25:50:0.8cm and 0.5cm) node(n2)
       {}
        (0.7,0) ++(50:0.8cm and 0.5cm) arc (50:80:0.8cm and 0.5cm) node(n3){}
        (0.7,0) ++(80:0.8cm and 0.5cm) arc (80:95:0.8cm and 0.5cm) node(n4){}
          (0.7,0) ++(95:0.8cm and 0.5cm) arc (95:120:0.8cm and 0.5cm)  node(n5) {} 
          (0.7,0) ++(120:0.8cm and 0.5cm) arc (120:155:0.8cm and 0.5cm) node(n6) {}
            (0.7,0) ++(155:0.8cm and 0.5cm) arc (155:360:0.8cm and 0.5cm)
             (n1.base) to  [out=240,in=250] (n4.base) 
             (n2.base) to  [out= 300,in= 260] (n3.base) 
     (n5.base) to  [out= 320,in= 300] (n6.base) 
         (1.5,0)--(2,0);
                   \filldraw [gray] (n1) circle [radius=1.5pt]; 
          \filldraw [gray]  (-0.1,0) circle [radius=1.5pt];
          \filldraw [gray] (n3) circle [radius=1.5pt]; 
           \filldraw [gray] (n5) circle [radius=1.5pt]; 
          }
          +
 \tikz[baseline=(vert_cent.base)]{
  \node (vert_cent) {\hspace{-13pt}$\phantom{-}$};
  \draw (-0.6,0)--(-0.1,0)
     (0.7,0) ++(0:0.8cm and 0.5cm) arc (0:25:0.8cm and 0.5cm) node (n1)
            {}
     (0.7,0) ++(25:0.8cm and 0.5cm) arc (25:60:0.8cm and 0.5cm) node(n2)
       {}
        (0.7,0) ++(60:0.8cm and 0.5cm) arc (60:85:0.8cm and 0.5cm) node(n3){}
        (0.7,0) ++(85:0.8cm and 0.5cm) arc (85:100:0.8cm and 0.5cm) node(n4){}
          (0.7,0) ++(100:0.8cm and 0.5cm) arc (100:130:0.8cm and 0.5cm)  node(n5) {} 
          (0.7,0) ++(130:0.8cm and 0.5cm) arc (130:155:0.8cm and 0.5cm) node(n6) {}
            (0.7,0) ++(155:0.8cm and 0.5cm) arc (155:360:0.8cm and 0.5cm)
             (n1.base) to  [out=240,in=240] (n2.base) 
             (n3.base) to  [out= 290,in= 300] (n6.base) 
     (n4.base) to  [out= 320,in= 270] (n5.base) 
         (1.5,0)--(2,0);
                   \filldraw [gray] (n1) circle [radius=1.5pt]; 
          \filldraw [gray]  (-0.1,0) circle [radius=1.5pt];
          \filldraw [gray] (n3) circle [radius=1.5pt]; 
           \filldraw [gray] (n5) circle [radius=1.5pt]; 
  }       \right ) 
  + c_{4D} \, 
  \tikz[baseline=(vert_cent.base)]{
  \node (vert_cent) {\hspace{-13pt}$\phantom{-}$};
  \draw (-0.6,0)--(-0.1,0)
     (0.7,0) ++(0:0.8cm and 0.5cm) arc (0:35:0.8cm and 0.5cm) node (n1)
            {}
     (0.7,0) ++(35:0.8cm and 0.5cm) arc (35:65:0.8cm and 0.5cm) node(n2)
       {}
        (0.7,0) ++(65:0.8cm and 0.5cm) arc (65:115:0.8cm and 0.5cm) node(n3){}
          (0.7,0) ++(115:0.8cm and 0.5cm) arc (115:145:0.8cm and 0.5cm)  node(n4) {} 
          (0.7,0) ++(145:0.8cm and 0.5cm) arc (145:240:0.8cm and 0.5cm)  node(n5){}
           (0.7,0) ++(240:0.8cm and 0.5cm) arc (240:300:0.8cm and 0.5cm) node(n6){}
          (0.7,0) ++(300:0.8cm and 0.5cm) arc (300:360:0.8cm and 0.5cm) 
       (n2.base) to  [out= 220,in= 320] (n3.base) 
       (n1.base) to  [out=220,in=320] (n4.base) 
         (n5.base) to  [out= 70,in= 110] (n6.base)
         (1.5,0)--(2,0);
                   \filldraw [gray] (n1) circle [radius=1.5pt]; 
          \filldraw [gray]  (-0.1,0) circle [radius=1.5pt];
          \filldraw [gray] (n3) circle [radius=1.5pt]; 
           \filldraw [gray] (n6) circle [radius=1.5pt]; 
          } 
          \nn \\
          &{}+ c_{4E} \,
          \tikz[baseline=(vert_cent.base)]{
  \node (vert_cent) {\hspace{-13pt}$\phantom{-}$};
  \draw (-0.6,0)--(-0.1,0)
   (0.7,0) ++(0:0.8cm and 0.5cm) arc (0:25:0.8cm and 0.5cm) node (n1) {}
     (0.7,0) ++(25:0.8cm and 0.5cm) arc (25:42.5:0.8cm and 0.5cm) node (n2)
            {}
     (0.7,0) ++(42.5:0.8cm and 0.5cm) arc (42.5:80:0.8cm and 0.5cm) node(n3)
       {}
        (0.7,0) ++(80:0.8cm and 0.5cm) arc (80:100:0.8cm and 0.5cm) node(n4){}
      (n2.base) to  [out=240,in=240] (n3.base) 
          (0.7,0) ++(100:0.8cm and 0.5cm) arc (100:137.5:0.8cm and 0.5cm)  node(n5){}
           (0.7,0) ++(137.5:0.8cm and 0.5cm) arc (137.5:155:0.8cm and 0.5cm)  node(n6) {} 
            (0.7,0) ++(155:0.8cm and 0.5cm) arc (155:360:0.8cm and 0.5cm) 
         (n4.base) to  [out= 290,in= 320] (n5.base) 
         (n1.base) to  [out= 220,in= 320] (n6.base) 
         (1.5,0)--(2,0);
                   \filldraw [gray] (n1) circle [radius=1.5pt]; 
          \filldraw [gray]  (-0.1,0) circle [radius=1.5pt];
          \filldraw [gray] (n3) circle [radius=1.5pt]; 
           \filldraw [gray] (n5) circle [radius=1.5pt];
          }
          + c_{4F} \,
          \tikz[baseline=(vert_cent.base)]{
  \node (vert_cent) {\hspace{-13pt}  $\phantom{-}$};
  \draw (-0.6,0)--(-0.1,0)
     (0.7,0) ++(0:0.8cm and 0.5cm) arc (0:35:0.8cm and 0.5cm) node (n1)
            {}
     (0.7,0) ++(35:0.8cm and 0.5cm) arc (35:65:0.8cm and 0.5cm) node(n2)
       {}
        (0.7,0) ++(65:0.8cm and 0.5cm) arc (65:115:0.8cm and 0.5cm) node(n3){}
          (0.7,0) ++(115:0.8cm and 0.5cm) arc (115:145:0.8cm and 0.5cm)  node(n4) {} 
          (0.7,0) ++(145:0.8cm and 0.5cm) arc (145:360:0.8cm and 0.5cm) 
       (n2.base) to  [out= 220,in= 320] (n3.base) 
        (0.7,0.7) ++(35:0.8cm and  -0.7cm) arc (35:65:0.8cm and -0.7cm) node(n5) {}
         (0.7,0.7) ++(65:0.8cm and  -0.7cm) arc (65:115:0.8cm and -0.7cm) node(n6) {}
          (0.7,0.7) ++(115:0.8cm and  -0.7cm) arc (115:145:0.8cm and -0.7cm) 
           (n5.base) to  [out= 140,in= 40] (n6.base)
         (1.5,0)--(2,0);
                   \filldraw [gray] (n1) circle [radius=1.5pt]; 
          \filldraw [gray]  (-0.1,0) circle [radius=1.5pt];
          \filldraw [gray] (n3) circle [radius=1.5pt]; 
           \filldraw [gray] (n6) circle [radius=1.5pt]; 
          }
          + c_{4G} \, 
          \tikz[baseline=(vert_cent.base)]{
  \node (vert_cent) {\hspace{-13pt}$\phantom{-}$};
  \draw (-0.6,0)--(-0.1,0)
     (0.7,0) ++(0:0.8cm and 0.5cm) arc (0:30:0.8cm and 0.5cm) node (n1)
            {}
     (0.7,0) ++(30:0.8cm and 0.5cm) arc (30:55:0.8cm and 0.5cm) node(n2)
       {}
        (0.7,0) ++(55:0.8cm and 0.5cm) arc (55:72.5:0.8cm and 0.5cm) node(n3){}
        (0.7,0) ++(72.5:0.8cm and 0.5cm) arc (72.5:107.5:0.8cm and 0.5cm) node(n4){}
          (0.7,0) ++(107.5:0.8cm and 0.5cm) arc (107.5:125:0.8cm and 0.5cm)  node(n5) {} 
          (0.7,0) ++(125:0.8cm and 0.5cm) arc (125:150:0.8cm and 0.5cm) node(n6) {}
            (0.7,0) ++(150:0.8cm and 0.5cm) arc (150:360:0.8cm and 0.5cm)
            (n1.base) to  [out=230,in=310] (n6.base) 
             (n2.base) to  [out= 280,in= 260] (n5.base) 
     (n3.base) to  [out= 280,in= 260] (n4.base) 
         (1.5,0)--(2,0);
                   \filldraw [gray] (n1) circle [radius=1.5pt]; 
          \filldraw [gray]  (-0.1,0) circle [radius=1.5pt];
          \filldraw [gray] (n3) circle [radius=1.5pt]; 
           \filldraw [gray] (n5) circle [radius=1.5pt]; 
          } \, ,
\label{gfour}
\end{align}
and, with non planar subgraphs,
\begin{align}
\gamma(Y,\bY)^{(4)}_2 ={}& c_{4H} \, \left (
 \tikz[baseline=(vert_cent.base)]{
  \node (vert_cent) {\hspace{-13pt}$\phantom{-}$};
  \draw (-0.6,0)--(-0.1,0)
     (0.7,0) ++(0:0.8cm and 0.5cm) arc (0:45:0.8cm and 0.5cm) node (n1)  {}
     (0.7,0) ++(45:0.8cm and 0.5cm) arc (45:95:0.8cm and 0.5cm) node(n2) {}
              (0.7,0) ++(95:0.8cm and 0.5cm) arc (95:115:0.8cm and 0.5cm)  node(n3) {} 
          (0.7,0) ++(115:0.8cm and 0.5cm) arc (115:155:0.8cm and 0.5cm) node(n4) {}
            (0.7,0) ++(155:0.8cm and 0.5cm) arc (155:265:0.8cm and 0.5cm)  node(n5) {}
             (0.7,0) ++(265:0.8cm and 0.5cm) arc (265:315:0.8cm and 0.5cm) node(n6) {}
              (0.7,0) ++(315:0.8cm and 0.5cm) arc (315:360:0.8cm and 0.5cm)
            (n1.base) to (1.03,0.05)
            (0.95,-0.05) to (n5.base)
           (n2.base) to (n6.base) 
     (n3.base) to  [out= 320,in= 300] (n4.base) 
         (1.5,0)--(2,0);
                   \filldraw [gray] (n1) circle [radius=1.5pt]; 
          \filldraw [gray]  (-0.1,0) circle [radius=1.5pt];
         \filldraw [gray] (n3) circle [radius=1.5pt]; 
         \filldraw [gray] (n6) circle [radius=1.5pt]; 
          }
          +
 \tikz[baseline=(vert_cent.base)]{
  \node (vert_cent) {\hspace{-13pt}$\phantom{-}$};
  \draw (-0.6,0)--(-0.1,0)
     (0.7,0) ++(0:0.8cm and 0.5cm) arc (0:25:0.8cm and 0.5cm) node (n1)
            {}
     (0.7,0) ++(25:0.8cm and 0.5cm) arc (25:65:0.8cm and 0.5cm) node(n2)
       {}
        (0.7,0) ++(65:0.8cm and 0.5cm) arc (65:85:0.8cm and 0.5cm) node(n3){}
        (0.7,0) ++(85:0.8cm and 0.5cm) arc (85:135:0.8cm and 0.5cm) node(n4){}
        (0.7,0) ++(135:0.8cm and 0.5cm) arc (135:225:0.8cm and 0.5cm)  node(n5) {} 
        (0.7,0) ++(225:0.8cm and 0.5cm) arc (225:275:0.8cm and 0.5cm) node(n6) {}
            (0.7,0) ++(275:0.8cm and 0.5cm) arc (275:360:0.8cm and 0.5cm)
             (n1.base) to  [out=240,in=240] (n2.base) 
             (n3.base) to  (0.43,0.05)
             (0.37,-0.05) to (n5.base) 
     (n4.base) to  (n6.base) 
         (1.5,0)--(2,0);
                   \filldraw [gray] (n1) circle [radius=1.5pt]; 
          \filldraw [gray]  (-0.1,0) circle [radius=1.5pt];
          \filldraw [gray] (n3) circle [radius=1.5pt]; 
           \filldraw [gray] (n6) circle [radius=1.5pt]; 
  }       
  \right ) \nn \\
  &{}+ c_{4I} \, 
  \tikz[baseline=(vert_cent.base)]{
  \node (vert_cent) {\hspace{-13pt}$\phantom{-}$};
  \draw (-0.6,0)--(-0.1,0)
     (0.7,0) ++(0:0.8cm and 0.5cm) arc (0:45:0.8cm and 0.5cm) node (n1)  {}
     (0.7,0) ++(45:0.8cm and 0.5cm) arc (45:73:0.8cm and 0.5cm) node(n5)  {}
     (0.7,0) ++(73:0.8cm and 0.5cm) arc (73:107:0.8cm and 0.5cm) node(n6)  {}
     (0.7,0) ++(107:0.8cm and 0.5cm) arc (107:135:0.8cm and 0.5cm) node(n2)  {}
        (0.7,0) ++(135:0.8cm and 0.5cm) arc (135:225:0.8cm and 0.5cm) node(n3){}
          (0.7,0) ++(225:0.8cm and 0.5cm) arc (225:315:0.8cm and 0.5cm)  node(n4) {} 
          (0.7,0) ++(315:0.8cm and 0.5cm) arc (315:360:0.8cm and 0.5cm) 
           (n1.base) to (0.76,0.05)
         (n2.base) to  (n4.base) 
         (0.65,-0.05) to (n3.base)
         (n5.base) to [out = 280,in = 260] (n6.base)
         (1.5,0)--(2,0);
                   \filldraw [gray] (n1) circle [radius=1.5pt]; 
          \filldraw [gray]  (-0.1,0) circle [radius=1.5pt];
          \filldraw [gray] (n4) circle [radius=1.5pt]; 
           \filldraw [gray] (n6) circle [radius=1.5pt];
          }
          + c_{4J} \, 
\tikz[baseline=(vert_cent.base)]{
  \node (vert_cent) {\hspace{-13pt}$\phantom{-}$};
  \draw (-0.6,0)--(-0.1,0)
     (0.7,0) ++(0:0.8cm and 0.5cm) arc (0:45:0.8cm and 0.5cm) node (n1)  {}
            (0.7,0) ++(135:0.8cm and 0.5cm) arc (135:360:0.8cm and 0.5cm) 
                (0.7,0.36) ++(0:0.55cm and 0.35cm) arc (0:50:0.55cm and 0.35cm) node (n2) {}
          (0.7,0.36) ++(50:0.55cm and 0.35cm) arc (50:130:0.55cm and 0.35cm)  node (n3)  {}
           (0.7,0.36) ++(130:0.55cm and 0.35cm) arc (130:230:0.55cm and 0.35cm) node (n4)  {}
            (0.7,0.36) ++(230:0.55cm and 0.35cm) arc (230:310:0.55cm and 0.35cm) node (n5)  {}
             (0.7,0.36) ++(310:0.55cm and 0.35cm) arc (310:360:0.55cm and 0.35cm) node (n6)  {}
             (n2.base) to (0.74,0.4)
             (0.62,0.3) to (n4.base)
             (n3.base) to (n5.base)
         (1.5,0)--(2,0);
                   \filldraw [gray] (n1) circle [radius=1.5pt]; 
         \filldraw [gray]  (-0.1,0) circle [radius=1.5pt];
         \filldraw [gray] (n3) circle [radius=1.5pt]; 
          \filldraw [gray] (n4) circle [radius=1.5pt];
          }  \, ,
\end{align}
and, with a new topology,
\be
\gamma(Y,\bY)^{(4)}_3 = c_{4K} \, 
\tikz[baseline=(vert_cent.base)]{
  \node (vert_cent) {\hspace{-13pt}$\phantom{-}$};
  \draw (-0.6,0)--(-0.1,0)
     (0.7,0) ++(0:0.8cm and 0.5cm) arc (0:50:0.8cm and 0.5cm) node (n1)  {}
     (0.7,0) ++(50:0.8cm and 0.5cm) arc (50:130:0.8cm and 0.5cm) node(n2) {}
              (0.7,0) ++(130:0.8cm and 0.5cm) arc (130:230:0.8cm and 0.5cm)  node(n3) {} 
          (0.7,0) ++(230:0.8cm and 0.5cm) arc (230:310:0.8cm and 0.5cm) node(n4) {}
            (0.7,0) ++(310:0.8cm and 0.5cm) arc (310:360:0.8cm and 0.5cm) 
                (n2.base) to [out = 330, in = 90] (0.45,0)
           (0.45,0) to [out = 270, in =30] (n3.base)
            (n1.base) to  [out= 220,in= 90]  (0.95,0)
            (0.95,0)  to [out = 270, in = 330] (n4.base) 
            (0.45,0) to (0.95,0) 
         (1.5,0)--(2,0);
                   \filldraw [gray] (n1) circle [radius=1.5pt]; 
          \filldraw [gray]  (-0.1,0) circle [radius=1.5pt];
         \filldraw [gray] (n4) circle [radius=1.5pt]; 
       \filldraw [gray] (0.45,0) circle [radius=1.5pt]; 
          } \, .
\ee

At this order \eqref{ctwo} has additional contributions of the general form
\begin{align}
c(Y,\bY)^{(3)} {} =  {}& \epsilon_{3A} \,
\tikz[baseline=(vert_cent.base)]{
  \node (vert_cent) {\hspace{-13pt}$\phantom{-}$};
  \draw (-0.4,0)--(0.1,0)
     (0.7,0) ++(0:0.6cm and 0.4cm) arc (0:35:0.6cm and 0.4cm) node (n1)
            {}
     (0.7,0) ++(35:0.6cm and 0.4cm) arc (35:75:0.6cm and 0.4cm) node(n2)
       {}
        (0.7,0) ++(75:0.6cm and 0.4cm) arc (75:105:0.6cm and 0.4cm) node(n3){}
      (n1.base) to  [out=270,in=240] (n2.base) 
          (0.7,0) ++(105:0.6cm and 0.4cm) arc (105:145:0.6cm and 0.4cm)  node(n4) {} 
          (0.7,0) ++(145:0.6cm and 0.4cm) arc (145:360:0.6cm and 0.4cm) 
       (n3.base) to  [out= 330,in= 300] (n4.base) 
         (1.3,0)--(1.8,0);
                   \filldraw [gray] (n1) circle [radius=1.5pt]; 
          \filldraw [gray]  (0.1,0) circle [radius=1.5pt];
          \filldraw [gray] (n3) circle [radius=1.5pt]; 
          }
+ \epsilon_{3B} \,
\tikz[baseline=(vert_cent.base)]{[scale=1.5]
  \node (vert_cent) {\hspace{-13pt}$\phantom{-}$};
  \draw (-0.4,0)--(0.1,0)
     (0.7,0) ++(0:0.6cm and 0.4cm) arc (0:50:0.6cm and 0.4cm) node (n1)
            {}
     (0.7,0) ++(50:0.6cm and 0.4cm) arc (50:130:0.6cm and 0.4cm) node(n2)
       {}
        (0.7,0) ++(130:0.6cm and 0.4cm) arc (130:230:0.6cm and 0.4cm) node(n3){}
          (n1.base) to[out=215,in=325] (n2.base) 
          (0.7,0) ++(230:0.6cm and 0.4cm) arc (230:310:0.6cm and 0.4cm)  node(n4) {} 
          (0.7,0) ++(310:0.6cm and 0.4cm) arc (310:360:0.6cm and 0.4cm) 
         (n3.base) to[out=-325,in=-215] (n4.base) 
         (1.3,0)--(1.8,0);
                   \filldraw [gray] (n1) circle [radius=1.5pt]; 
          \filldraw [gray]  (0.1,0) circle [radius=1.5pt];
          \filldraw [gray] (n4) circle [radius=1.5pt]; 
          } 
{}+ \epsilon_{3C} \, 
\tikz[baseline=(vert_cent.base)]{
  \node (vert_cent) {\hspace{-13pt}$\phantom{-}$};
  \draw (-0.4,0)--(0.1,0)
     (0.7,0) ++(0:0.6cm and 0.4cm) arc (0:35:0.6cm and 0.4cm) node (n1)
            {}
     (0.7,0) ++(35:0.6cm and 0.4cm) arc (35:65:0.6cm and 0.4cm) node(n2)
       {}
        (0.7,0) ++(65:0.6cm and 0.4cm) arc (65:115:0.6cm and 0.4cm) node(n3){}
          (0.7,0) ++(115:0.6cm and 0.4cm) arc (115:145:0.6cm and 0.4cm)  node(n4) {} 
          (0.7,0) ++(145:0.6cm and 0.4cm) arc (145:360:0.6cm and 0.4cm) 
       (n2.base) to  [out= 220,in= 320] (n3.base) 
       (n1.base) to  [out=220,in=320] (n4.base) 
         (1.3,0)--(1.8,0);
                   \filldraw [gray] (n1) circle [radius=1.5pt]; 
          \filldraw [gray]  (0.1,0) circle [radius=1.5pt];
          \filldraw [gray] (n3) circle [radius=1.5pt]; 
          } \nn \\
&{} + \epsilon_{3D} \,
\tikz[baseline=(vert_cent.base)]{
  \node (vert_cent) {\hspace{-13pt}$\phantom{-}$};
  \draw (-0.4,0)--(0.1,0)
     (0.7,0) ++(0:0.6cm and 0.4cm) arc (0:50:0.6cm and 0.4cm) node (n1)
            {}
     (0.7,0) ++(50:0.6cm and 0.4cm) arc (50:130:0.6cm and 0.4cm) node(n2)
       {}
        (0.7,0) ++(130:0.6cm and 0.4cm) arc (130:230:0.6cm and 0.4cm) node(n3){}
          (0.7,0) ++(230:0.6cm and 0.4cm) arc (230:310:0.6cm and 0.4cm)  node(n4) {} 
          (0.7,0) ++(310:0.6cm and 0.4cm) arc (310:360:0.6cm and 0.4cm) 
           (n1.base) to (0.75,0.05)
         (n2.base) to  (n4.base) 
         (0.65,-0.05) to (n3.base)
         (1.3,0)--(1.8,0);
                   \filldraw [gray] (n1) circle [radius=1.5pt]; 
          \filldraw [gray]  (0.1,0) circle [radius=1.5pt];
          \filldraw [gray] (n4) circle [radius=1.5pt]; 
          }
       \,    ,
          \label{cthree}
\end{align}
while \eqref{cthreeb} is extended by
\begin{align}
c'(Y',\bY')^{(3)}  ={}&  - ( \epsilon_{3A} - 4 \, \epsilon_1 \epsilon_2 + 8 \, \epsilon_1{\!}^3 ) \,
\tikz[baseline=(vert_cent.base)]{
  \node (vert_cent) {\hspace{-13pt}$\phantom{-}$};
  \draw (-0.4,0)--(0.1,0)
     (0.7,0) ++(0:0.6cm and 0.4cm) arc (0:35:0.6cm and 0.4cm) node (n1)
            {}
     (0.7,0) ++(35:0.6cm and 0.4cm) arc (35:75:0.6cm and 0.4cm) node(n2)
       {}
        (0.7,0) ++(75:0.6cm and 0.4cm) arc (75:105:0.6cm and 0.4cm) node(n3){}
      (n1.base) to  [out=270,in=240] (n2.base) 
          (0.7,0) ++(105:0.6cm and 0.4cm) arc (105:145:0.6cm and 0.4cm)  node(n4) {} 
          (0.7,0) ++(145:0.6cm and 0.4cm) arc (145:360:0.6cm and 0.4cm) 
       (n3.base) to  [out= 330,in= 300] (n4.base) 
         (1.3,0)--(1.8,0);
                   \filldraw [gray] (n1) circle [radius=1.5pt]; 
          \filldraw [gray]  (0.1,0) circle [radius=1.5pt];
          \filldraw [gray] (n3) circle [radius=1.5pt]; 
          }
- ( \epsilon_{3B}  - 2 \, \epsilon_1 \epsilon_2 + 4 \, \epsilon_1{\!}^3 ) \,
\tikz[baseline=(vert_cent.base)]{[scale=1.5]
  \node (vert_cent) {\hspace{-13pt}$\phantom{-}$};
  \draw (-0.4,0)--(0.1,0)
     (0.7,0) ++(0:0.6cm and 0.4cm) arc (0:50:0.6cm and 0.4cm) node (n1)
            {}
     (0.7,0) ++(50:0.6cm and 0.4cm) arc (50:130:0.6cm and 0.4cm) node(n2)
       {}
        (0.7,0) ++(130:0.6cm and 0.4cm) arc (130:230:0.6cm and 0.4cm) node(n3){}
          (n1.base) to[out=215,in=325] (n2.base) 
          (0.7,0) ++(230:0.6cm and 0.4cm) arc (230:310:0.6cm and 0.4cm)  node(n4) {} 
          (0.7,0) ++(310:0.6cm and 0.4cm) arc (310:360:0.6cm and 0.4cm) 
         (n3.base) to[out=-325,in=-215] (n4.base) 
         (1.3,0)--(1.8,0);
                   \filldraw [gray] (n1) circle [radius=1.5pt]; 
          \filldraw [gray]  (0.1,0) circle [radius=1.5pt];
          \filldraw [gray] (n4) circle [radius=1.5pt]; 
          } 
          \nn \\ &
{} - (\epsilon_{3C}  - 8 \, \epsilon_1 \epsilon_2 + 16 \, \epsilon_1{\!}^3 ) \,
\tikz[baseline=(vert_cent.base)]{
  \node (vert_cent) {\hspace{-13pt}$\phantom{-}$};
  \draw (-0.4,0)--(0.1,0)
     (0.7,0) ++(0:0.6cm and 0.4cm) arc (0:35:0.6cm and 0.4cm) node (n1)
            {}
     (0.7,0) ++(35:0.6cm and 0.4cm) arc (35:65:0.6cm and 0.4cm) node(n2)
       {}
        (0.7,0) ++(65:0.6cm and 0.4cm) arc (65:115:0.6cm and 0.4cm) node(n3){}
          (0.7,0) ++(115:0.6cm and 0.4cm) arc (115:145:0.6cm and 0.4cm)  node(n4) {} 
          (0.7,0) ++(145:0.6cm and 0.4cm) arc (145:360:0.6cm and 0.4cm) 
       (n2.base) to  [out= 220,in= 320] (n3.base) 
       (n1.base) to  [out=220,in=320] (n4.base) 
         (1.3,0)--(1.8,0);
                   \filldraw [gray] (n1) circle [radius=1.5pt]; 
          \filldraw [gray]  (0.1,0) circle [radius=1.5pt];
          \filldraw [gray] (n3) circle [radius=1.5pt]; 
          }
- \epsilon_{3D} \,
\tikz[baseline=(vert_cent.base)]{
  \node (vert_cent) {\hspace{-13pt}$\phantom{-}$};
  \draw (-0.4,0)--(0.1,0)
     (0.7,0) ++(0:0.6cm and 0.4cm) arc (0:50:0.6cm and 0.4cm) node (n1)
            {}
     (0.7,0) ++(50:0.6cm and 0.4cm) arc (50:130:0.6cm and 0.4cm) node(n2)
       {}
        (0.7,0) ++(130:0.6cm and 0.4cm) arc (130:230:0.6cm and 0.4cm) node(n3){}
          (0.7,0) ++(230:0.6cm and 0.4cm) arc (230:310:0.6cm and 0.4cm)  node(n4) {} 
          (0.7,0) ++(310:0.6cm and 0.4cm) arc (310:360:0.6cm and 0.4cm) 
           (n1.base) to (0.75,0.05)
         (n2.base) to  (n4.base) 
         (0.65,-0.05) to (n3.base)
         (1.3,0)--(1.8,0);
                   \filldraw [gray] (n1) circle [radius=1.5pt]; 
          \filldraw [gray]  (0.1,0) circle [radius=1.5pt];
          \filldraw [gray] (n4) circle [radius=1.5pt]; 
          }
       \,    .
          \label{cfourb}
\end{align}
In this case
\begin{align}
c_\beta(Y,\bY)^{(4)}_1 ={}& 6\, c_1 \epsilon_{3A} \, 
\tikz[baseline=(vert_cent.base)]{
  \node (vert_cent) {\hspace{-13pt}$\phantom{-}$};
  \draw (-0.6,0)--(-0.1,0)
     (0.7,0) ++(0:0.8cm and 0.5cm) arc (0:25:0.8cm and 0.5cm) node (n1)
            {}
     (0.7,0) ++(25:0.8cm and 0.5cm) arc (25:60:0.8cm and 0.5cm) node(n2)
       {}
        (0.7,0) ++(60:0.8cm and 0.5cm) arc (60:72.5:0.8cm and 0.5cm) node(n3){}
      (n1.base) to  [out=270,in=250] (n2.base) 
        (0.7,0) ++(72.5:0.8cm and 0.5cm) arc (72.5:107.5:0.8cm and 0.5cm) node(n4){}
          (0.7,0) ++(107.5:0.8cm and 0.5cm) arc (107.5:120:0.8cm and 0.5cm)  node(n5) {} 
          (0.7,0) ++(120:0.8cm and 0.5cm) arc (120:155:0.8cm and 0.5cm) node(n6) {}
            (0.7,0) ++(155:0.8cm and 0.5cm) arc (155:360:0.8cm and 0.5cm)
             (n3.base) to  [out= 280,in= 260] (n4.base) 
     (n5.base) to  [out= 320,in= 300] (n6.base) 
         (1.5,0)--(2,0);
                   \filldraw [gray] (n1) circle [radius=1.5pt]; 
          \filldraw [gray]  (-0.1,0) circle [radius=1.5pt];
          \filldraw [gray] (n3) circle [radius=1.5pt]; 
           \filldraw [gray] (n5) circle [radius=1.5pt]; 
          }
+ 2\, c_1 (\epsilon_{3A} + 4\, \epsilon_{3B}) \, 
\tikz[baseline=(vert_cent.base)]{
  \node (vert_cent) {\hspace{-13pt}$\phantom{-}$};
  \draw (-0.6,0)--(-0.1,0)
     (0.7,0) ++(0:0.8cm and 0.5cm) arc (0:35:0.8cm and 0.5cm) node (n1)
            {}
     (0.7,0) ++(35:0.8cm and 0.5cm) arc (35:75:0.8cm and 0.5cm) node(n2)
       {}
        (0.7,0) ++(75:0.8cm and 0.5cm) arc (75:105:0.8cm and 0.5cm) node(n3){}
      (n1.base) to  [out=270,in=240] (n2.base) 
          (0.7,0) ++(105:0.8cm and 0.5cm) arc (105:145:0.8cm and 0.5cm)  node(n4) {} 
          (0.7,0) ++(145:0.8cm and 0.5cm) arc (145:240:0.8cm and 0.5cm) node(n5){}
          (0.7,0) ++(240:0.8cm and 0.5cm) arc (240:300:0.8cm and 0.5cm) node(n6){}
          (0.7,0) ++(300:0.8cm and 0.5cm) arc (300:360:0.8cm and 0.5cm) 
       (n3.base) to  [out= 280,in= 320] (n4.base) 
       (n5.base) to  [out= 80,in= 100] (n6.base) 
         (1.5,0)--(2,0);
                   \filldraw [gray] (n1) circle [radius=1.5pt]; 
          \filldraw [gray]  (-0.1,0) circle [radius=1.5pt];
          \filldraw [gray] (n3) circle [radius=1.5pt]; 
           \filldraw [gray] (n6) circle [radius=1.5pt];
           }
           \nn \\
           &{}+ 2 ( 2\, c_1 \, \epsilon_{3A} + c_1 \,\epsilon_{3C} + c_2\,  \epsilon_2)  \left (
           \tikz[baseline=(vert_cent.base)]{
  \node (vert_cent) {\hspace{-13pt}$\phantom{-}$};
  \draw (-0.6,0)--(-0.1,0)
     (0.7,0) ++(0:0.8cm and 0.5cm) arc (0:25:0.8cm and 0.5cm) node (n1)
            {}
     (0.7,0) ++(25:0.8cm and 0.5cm) arc (25:50:0.8cm and 0.5cm) node(n2)
       {}
        (0.7,0) ++(50:0.8cm and 0.5cm) arc (50:80:0.8cm and 0.5cm) node(n3){}
        (0.7,0) ++(80:0.8cm and 0.5cm) arc (80:95:0.8cm and 0.5cm) node(n4){}
          (0.7,0) ++(95:0.8cm and 0.5cm) arc (95:120:0.8cm and 0.5cm)  node(n5) {} 
          (0.7,0) ++(120:0.8cm and 0.5cm) arc (120:155:0.8cm and 0.5cm) node(n6) {}
            (0.7,0) ++(155:0.8cm and 0.5cm) arc (155:360:0.8cm and 0.5cm)
             (n1.base) to  [out=240,in=250] (n4.base) 
             (n2.base) to  [out= 300,in= 260] (n3.base) 
     (n5.base) to  [out= 320,in= 300] (n6.base) 
         (1.5,0)--(2,0);
                   \filldraw [gray] (n1) circle [radius=1.5pt]; 
          \filldraw [gray]  (-0.1,0) circle [radius=1.5pt];
          \filldraw [gray] (n3) circle [radius=1.5pt]; 
           \filldraw [gray] (n5) circle [radius=1.5pt]; 
          }
          +
 \tikz[baseline=(vert_cent.base)]{
  \node (vert_cent) {\hspace{-13pt}$\phantom{-}$};
  \draw (-0.6,0)--(-0.1,0)
     (0.7,0) ++(0:0.8cm and 0.5cm) arc (0:25:0.8cm and 0.5cm) node (n1)
            {}
     (0.7,0) ++(25:0.8cm and 0.5cm) arc (25:60:0.8cm and 0.5cm) node(n2)
       {}
        (0.7,0) ++(60:0.8cm and 0.5cm) arc (60:85:0.8cm and 0.5cm) node(n3){}
        (0.7,0) ++(85:0.8cm and 0.5cm) arc (85:100:0.8cm and 0.5cm) node(n4){}
          (0.7,0) ++(100:0.8cm and 0.5cm) arc (100:130:0.8cm and 0.5cm)  node(n5) {} 
          (0.7,0) ++(130:0.8cm and 0.5cm) arc (130:155:0.8cm and 0.5cm) node(n6) {}
            (0.7,0) ++(155:0.8cm and 0.5cm) arc (155:360:0.8cm and 0.5cm)
             (n1.base) to  [out=240,in=240] (n2.base) 
             (n3.base) to  [out= 290,in= 300] (n6.base) 
     (n4.base) to  [out= 320,in= 270] (n5.base) 
         (1.5,0)--(2,0);
                   \filldraw [gray] (n1) circle [radius=1.5pt]; 
          \filldraw [gray]  (-0.1,0) circle [radius=1.5pt];
          \filldraw [gray] (n3) circle [radius=1.5pt]; 
           \filldraw [gray] (n5) circle [radius=1.5pt]; 
  }       \right ) \nn \\
  &{} + 2(4 \, c_1 \, \epsilon_{3B} + c_1 \,\epsilon_{3C} + c_2  \, \epsilon_2) \, 
  \tikz[baseline=(vert_cent.base)]{
  \node (vert_cent) {\hspace{-13pt}$\phantom{-}$};
  \draw (-0.6,0)--(-0.1,0)
     (0.7,0) ++(0:0.8cm and 0.5cm) arc (0:35:0.8cm and 0.5cm) node (n1)
            {}
     (0.7,0) ++(35:0.8cm and 0.5cm) arc (35:65:0.8cm and 0.5cm) node(n2)
       {}
        (0.7,0) ++(65:0.8cm and 0.5cm) arc (65:115:0.8cm and 0.5cm) node(n3){}
          (0.7,0) ++(115:0.8cm and 0.5cm) arc (115:145:0.8cm and 0.5cm)  node(n4) {} 
          (0.7,0) ++(145:0.8cm and 0.5cm) arc (145:240:0.8cm and 0.5cm)  node(n5){}
           (0.7,0) ++(240:0.8cm and 0.5cm) arc (240:300:0.8cm and 0.5cm) node(n6){}
          (0.7,0) ++(300:0.8cm and 0.5cm) arc (300:360:0.8cm and 0.5cm) 
       (n2.base) to  [out= 220,in= 320] (n3.base) 
       (n1.base) to  [out=220,in=320] (n4.base) 
         (n5.base) to  [out= 70,in= 110] (n6.base)
         (1.5,0)--(2,0);
                   \filldraw [gray] (n1) circle [radius=1.5pt]; 
          \filldraw [gray]  (-0.1,0) circle [radius=1.5pt];
          \filldraw [gray] (n3) circle [radius=1.5pt]; 
           \filldraw [gray] (n6) circle [radius=1.5pt]; 
          } 
          + 4( c_1 \, \epsilon_{3C} +  c_{3A}\,  \epsilon_1)  \,
          \tikz[baseline=(vert_cent.base)]{
  \node (vert_cent) {\hspace{-13pt}$\phantom{-}$};
  \draw (-0.6,0)--(-0.1,0)
   (0.7,0) ++(0:0.8cm and 0.5cm) arc (0:25:0.8cm and 0.5cm) node (n1) {}
     (0.7,0) ++(25:0.8cm and 0.5cm) arc (25:42.5:0.8cm and 0.5cm) node (n2)
            {}
     (0.7,0) ++(42.5:0.8cm and 0.5cm) arc (42.5:80:0.8cm and 0.5cm) node(n3)
       {}
        (0.7,0) ++(80:0.8cm and 0.5cm) arc (80:100:0.8cm and 0.5cm) node(n4){}
      (n2.base) to  [out=240,in=240] (n3.base) 
          (0.7,0) ++(100:0.8cm and 0.5cm) arc (100:137.5:0.8cm and 0.5cm)  node(n5){}
           (0.7,0) ++(137.5:0.8cm and 0.5cm) arc (137.5:155:0.8cm and 0.5cm)  node(n6) {} 
            (0.7,0) ++(155:0.8cm and 0.5cm) arc (155:360:0.8cm and 0.5cm) 
         (n4.base) to  [out= 290,in= 320] (n5.base) 
         (n1.base) to  [out= 220,in= 320] (n6.base) 
         (1.5,0)--(2,0);
                   \filldraw [gray] (n1) circle [radius=1.5pt]; 
          \filldraw [gray]  (-0.1,0) circle [radius=1.5pt];
          \filldraw [gray] (n3) circle [radius=1.5pt]; 
           \filldraw [gray] (n5) circle [radius=1.5pt];
          } 
\nn \\
&{} + 2 ( c_1 \, \epsilon_{3C} + 2 \, c_{3B}\,  \epsilon_1)  \,
          \tikz[baseline=(vert_cent.base)]{
  \node (vert_cent) {\hspace{-13pt}  $\phantom{-}$};
  \draw (-0.6,0)--(-0.1,0)
     (0.7,0) ++(0:0.8cm and 0.5cm) arc (0:35:0.8cm and 0.5cm) node (n1)
            {}
     (0.7,0) ++(35:0.8cm and 0.5cm) arc (35:65:0.8cm and 0.5cm) node(n2)
       {}
        (0.7,0) ++(65:0.8cm and 0.5cm) arc (65:115:0.8cm and 0.5cm) node(n3){}
          (0.7,0) ++(115:0.8cm and 0.5cm) arc (115:145:0.8cm and 0.5cm)  node(n4) {} 
          (0.7,0) ++(145:0.8cm and 0.5cm) arc (145:360:0.8cm and 0.5cm) 
       (n2.base) to  [out= 220,in= 320] (n3.base) 
        (0.7,0.7) ++(35:0.8cm and  -0.7cm) arc (35:65:0.8cm and -0.7cm) node(n5) {}
         (0.7,0.7) ++(65:0.8cm and  -0.7cm) arc (65:115:0.8cm and -0.7cm) node(n6) {}
          (0.7,0.7) ++(115:0.8cm and  -0.7cm) arc (115:145:0.8cm and -0.7cm) 
           (n5.base) to  [out= 140,in= 40] (n6.base)
         (1.5,0)--(2,0);
                   \filldraw [gray] (n1) circle [radius=1.5pt]; 
          \filldraw [gray]  (-0.1,0) circle [radius=1.5pt];
          \filldraw [gray] (n3) circle [radius=1.5pt]; 
           \filldraw [gray] (n6) circle [radius=1.5pt]; 
          }
          + 4 ( c_1\, \epsilon_{3C} +c_2   \, \epsilon_2 +  c_{3C}\,  \epsilon_1 )  
          \tikz[baseline=(vert_cent.base)]{
  \node (vert_cent) {\hspace{-13pt}$\phantom{-}$};
  \draw (-0.6,0)--(-0.1,0)
     (0.7,0) ++(0:0.8cm and 0.5cm) arc (0:30:0.8cm and 0.5cm) node (n1)
            {}
     (0.7,0) ++(30:0.8cm and 0.5cm) arc (30:55:0.8cm and 0.5cm) node(n2)
       {}
        (0.7,0) ++(55:0.8cm and 0.5cm) arc (55:72.5:0.8cm and 0.5cm) node(n3){}
        (0.7,0) ++(72.5:0.8cm and 0.5cm) arc (72.5:107.5:0.8cm and 0.5cm) node(n4){}
          (0.7,0) ++(107.5:0.8cm and 0.5cm) arc (107.5:125:0.8cm and 0.5cm)  node(n5) {} 
          (0.7,0) ++(125:0.8cm and 0.5cm) arc (125:150:0.8cm and 0.5cm) node(n6) {}
            (0.7,0) ++(150:0.8cm and 0.5cm) arc (150:360:0.8cm and 0.5cm)
            (n1.base) to  [out=230,in=310] (n6.base) 
             (n2.base) to  [out= 280,in= 260] (n5.base) 
     (n3.base) to  [out= 280,in= 260] (n4.base) 
         (1.5,0)--(2,0);
                   \filldraw [gray] (n1) circle [radius=1.5pt]; 
          \filldraw [gray]  (-0.1,0) circle [radius=1.5pt];
          \filldraw [gray] (n3) circle [radius=1.5pt]; 
           \filldraw [gray] (n5) circle [radius=1.5pt]; 
          } \, ,
\end{align}
and the non planar contributions are
\begin{align}
c_\beta(Y,\bY)_2^{(4)} ={}& 4\, c_1\, \epsilon_{3D} \, \left (
 \tikz[baseline=(vert_cent.base)]{
  \node (vert_cent) {\hspace{-13pt}$\phantom{-}$};
  \draw (-0.6,0)--(-0.1,0)
     (0.7,0) ++(0:0.8cm and 0.5cm) arc (0:45:0.8cm and 0.5cm) node (n1)  {}
     (0.7,0) ++(45:0.8cm and 0.5cm) arc (45:95:0.8cm and 0.5cm) node(n2) {}
              (0.7,0) ++(95:0.8cm and 0.5cm) arc (95:120:0.8cm and 0.5cm)  node(n3) {} 
          (0.7,0) ++(120:0.8cm and 0.5cm) arc (120:155:0.8cm and 0.5cm) node(n4) {}
            (0.7,0) ++(155:0.8cm and 0.5cm) arc (155:265:0.8cm and 0.5cm)  node(n5) {}
             (0.7,0) ++(265:0.8cm and 0.5cm) arc (265:315:0.8cm and 0.5cm) node(n6) {}
              (0.7,0) ++(315:0.8cm and 0.5cm) arc (315:360:0.8cm and 0.5cm)
            (n1.base) to (1.03,0.05)
            (0.95,-0.05) to (n5.base)
           (n2.base) to (n6.base) 
     (n3.base) to  [out= 320,in= 300] (n4.base) 
         (1.5,0)--(2,0);
                   \filldraw [gray] (n1) circle [radius=1.5pt]; 
          \filldraw [gray]  (-0.1,0) circle [radius=1.5pt];
         \filldraw [gray] (n3) circle [radius=1.5pt]; 
         \filldraw [gray] (n6) circle [radius=1.5pt]; 
          }
          +
 \tikz[baseline=(vert_cent.base)]{
  \node (vert_cent) {\hspace{-13pt}$\phantom{-}$};
  \draw (-0.6,0)--(-0.1,0)
     (0.7,0) ++(0:0.8cm and 0.5cm) arc (0:25:0.8cm and 0.5cm) node (n1)
            {}
     (0.7,0) ++(25:0.8cm and 0.5cm) arc (25:60:0.8cm and 0.5cm) node(n2)
       {}
        (0.7,0) ++(60:0.8cm and 0.5cm) arc (60:85:0.8cm and 0.5cm) node(n3){}
        (0.7,0) ++(85:0.8cm and 0.5cm) arc (85:135:0.8cm and 0.5cm) node(n4){}
        (0.7,0) ++(135:0.8cm and 0.5cm) arc (135:225:0.8cm and 0.5cm)  node(n5) {} 
        (0.7,0) ++(225:0.8cm and 0.5cm) arc (225:275:0.8cm and 0.5cm) node(n6) {}
            (0.7,0) ++(275:0.8cm and 0.5cm) arc (275:360:0.8cm and 0.5cm)
             (n1.base) to  [out=240,in=240] (n2.base) 
             (n3.base) to  (0.43,0.05)
             (0.37,-0.05) to (n5.base) 
     (n4.base) to  (n6.base) 
         (1.5,0)--(2,0);
                   \filldraw [gray] (n1) circle [radius=1.5pt]; 
          \filldraw [gray]  (-0.1,0) circle [radius=1.5pt];
          \filldraw [gray] (n3) circle [radius=1.5pt]; 
           \filldraw [gray] (n6) circle [radius=1.5pt]; 
  }       
   + 2 \, 
  \tikz[baseline=(vert_cent.base)]{
  \node (vert_cent) {\hspace{-13pt}$\phantom{-}$};
  \draw (-0.6,0)--(-0.1,0)
     (0.7,0) ++(0:0.8cm and 0.5cm) arc (0:45:0.8cm and 0.5cm) node (n1)  {}
     (0.7,0) ++(45:0.8cm and 0.5cm) arc (45:73:0.8cm and 0.5cm) node(n5)  {}
     (0.7,0) ++(73:0.8cm and 0.5cm) arc (73:107:0.8cm and 0.5cm) node(n6)  {}
     (0.7,0) ++(107:0.8cm and 0.5cm) arc (107:135:0.8cm and 0.5cm) node(n2)  {}
        (0.7,0) ++(135:0.8cm and 0.5cm) arc (135:225:0.8cm and 0.5cm) node(n3){}
          (0.7,0) ++(225:0.8cm and 0.5cm) arc (225:315:0.8cm and 0.5cm)  node(n4) {} 
          (0.7,0) ++(315:0.8cm and 0.5cm) arc (315:360:0.8cm and 0.5cm) 
           (n1.base) to (0.76,0.05)
         (n2.base) to  (n4.base) 
         (0.65,-0.05) to (n3.base)
         (n5.base) to [out = 280,in = 260] (n6.base)
         (1.5,0)--(2,0);
                   \filldraw [gray] (n1) circle [radius=1.5pt]; 
          \filldraw [gray]  (-0.1,0) circle [radius=1.5pt];
          \filldraw [gray] (n4) circle [radius=1.5pt]; 
           \filldraw [gray] (n6) circle [radius=1.5pt];
          }
           \right ) \nn \\
  &{}        + 4\, c_{3D} \, \epsilon_1 \, 
\tikz[baseline=(vert_cent.base)]{
  \node (vert_cent) {\hspace{-13pt}$\phantom{-}$};
  \draw (-0.6,0)--(-0.1,0)
     (0.7,0) ++(0:0.8cm and 0.5cm) arc (0:45:0.8cm and 0.5cm) node (n1)  {}
            (0.7,0) ++(135:0.8cm and 0.5cm) arc (135:360:0.8cm and 0.5cm) 
                (0.7,0.36) ++(0:0.55cm and 0.35cm) arc (0:50:0.55cm and 0.35cm) node (n2) {}
          (0.7,0.36) ++(50:0.55cm and 0.35cm) arc (50:130:0.55cm and 0.35cm)  node (n3)  {}
           (0.7,0.36) ++(130:0.55cm and 0.35cm) arc (130:230:0.55cm and 0.35cm) node (n4)  {}
            (0.7,0.36) ++(230:0.55cm and 0.35cm) arc (230:310:0.55cm and 0.35cm) node (n5)  {}
             (0.7,0.36) ++(310:0.55cm and 0.35cm) arc (310:360:0.55cm and 0.35cm) node (n6)  {}
             (n2.base) to (0.74,0.4)
             (0.62,0.3) to (n4.base)
             (n3.base) to (n5.base)
         (1.5,0)--(2,0);
                   \filldraw [gray] (n1) circle [radius=1.5pt]; 
         \filldraw [gray]  (-0.1,0) circle [radius=1.5pt];
         \filldraw [gray] (n3) circle [radius=1.5pt]; 
          \filldraw [gray] (n4) circle [radius=1.5pt];
          }  \, .
\end{align}
Transforming $(Y,\bY) \to (Y',\bY')$ now gives $\gamma'(Y',\bY')^{(4)}$ of
the form \eqref{gfour} with
\begin{align}
c'{\!}_{4A} = {}& c_{4A} + 32\, c_1 \, \epsilon_1{\!}^3 -24  \, c_1\, \epsilon_1 \epsilon_2 +
 12 \, c_2\,  \epsilon_1{\!}^2 - 6\, c_{3A}\, \epsilon_1 + 6 \, c_1\, \epsilon_{3A} \, , \nn \\
 c'{\!}_{4B} = {}& c_{4B} + 32\, c_1 \, \epsilon_1{\!}^3 - 24  \, c_1\, \epsilon_1 \epsilon_2 +
 12 \, c_2\,  \epsilon_1{\!}^2 - 2( c_{3A} +4\, c_{3B} )\epsilon_1 + 
 2\, c_1( \epsilon_{3A} + 4\, \epsilon_{3B} )  \, , \nn \\
  c'{\!}_{4C} = {}& c_{4C} + 32\, c_1 \, \epsilon_1{\!}^3 - 24  \, c_1\, \epsilon_1 \epsilon_2 +
 8 \, c_2\,  \epsilon_1{\!}^2 - 2( 2\, c_{3A} + c_{3C} )\epsilon_1 + 
 2\, c_1(2\,  \epsilon_{3A} +  \epsilon_{3C} )  \, , \nn \\
  c'{\!}_{4D} = {}& c_{4D} + 32\, c_1 \, \epsilon_1{\!}^3 - 24  \, c_1\, \epsilon_1 \epsilon_2 +
 8 \, c_2\,  \epsilon_1{\!}^2 - 2( 4\, c_{3B} + c_{3C} )\epsilon_1 + 
 2\, c_1( 4\, \epsilon_{3B} +  \epsilon_{3C} )  \, , \nn \\
  c'{\!}_{4E} = {}& c_{4E} 
 -8 \, c_2\,  \epsilon_1{\!}^2  + 4 ( c_{3A}  - c_{3C} )\epsilon_1 - 
 4\,  c_1( \epsilon_{3A}  - \epsilon_{3C} )  \, , \nn \\
   c'{\!}_{4F} = {}& c_{4F} 
 -4  \, c_2\,  \epsilon_1{\!}^2  + 2 ( 2 c_{3B}  - c_{3C} )\epsilon_1 - 
 2\,  c_1(2\,  \epsilon_{3B}  - \epsilon_{3C} )  \, , \nn \\
    c'{\!}_{4G} = {}& c_{4G} \, ,
\label{tran4}
\end{align}
and
\be
c'{\!}_{4H} = c_{4H}  + Y \, , \quad c'{\!}_{4I} = c_{4I}  + 2\, Y \, , \quad c'{\!}_{4J} = c_{4J}  - Y  \, ,
\quad Y = 4 ( c_1 \, \epsilon_{3D} - c_{3D} \, \epsilon_1 ) \, .
\label{tran4n}
\ee

In consequence of \eqref{tran4} 4 loop scheme invariants are given, for the planar graphs, by
\be
c_{4A} - c_{4B} - c_{4C} + c_{4D} \, , \quad 2\, c_{4A} - 2\, c_{4C} + c_{4E} \, , \quad
c_{4A} + c_{4B} - 2 \, c_{4C} + 2\, c_{4F} \, , \quad c_{4G} \, ,
\ee 
and for the non planar ones from \eqref{tran4n}
\be
2 \, c_{4H} - c_{4I} \, , \qquad c_{4H} + c_{4J} \, .
\ee

The results of \cite{Jack4}\footnote{With our notation their individual results calculated
with a minimal subtraction scheme, if not quoted in the text, are
$c_{4A}=c_{4B} = \tfrac18(2\zeta(3)-1), \, c_{4C}=c_{4D} = \tfrac13, \, c_{4E} = \tfrac{1}{12}
( 5 - 6 \zeta(3), \, c_{4F} = \tfrac{5}{24}$ and for the non planar graphs $ 2 c_{4H} = c_{4I} = - \half 
(6 \zeta(3) - 3 \zeta(4)) , \, c_{4J} = - \tfrac34( 6 \zeta(3)- 3 \zeta(4))$. Corresponding results for 
$c_{4B}+c_{4D}$ and $c_{4A} + 2c_{4C}+c_{4E}+c_{4F}+c_{4G}+c_{4J}$ as well as
$c_{4I}=2c_{4H}, \, c_{4K}$ can be read off from \cite{Avdeev}.} give scheme invariants 
which may be expressed in terms of a basis given by
\begin{align}
& c_{4A} - c_{4B} - c_{4C} + c_{4D}  = 0 \, , \nn \\
&  2\, c_{4A} - 2\, c_{4C} + c_{4E} = - \half  \, , \quad
c_{4A} + c_{4B} - 2 \, c_{4C} + 2\, c_{4F} = \half \zeta(3) - \half  \, , \quad c_{4G} = - \tfrac52 \, , \nn \\
& 2 \, c_{4H} - c_{4I} =0 \, , \qquad c_{4H} + c_{4J} = -3 \, \zeta(3) \, , \qquad
c_{4K} = - 10 \, \zeta(5) \, .
\label{ResJ}
\end{align}
Using the freedom allowed by \eqref{tran4} and the results in \cite{Jack4} there is a minimal
scheme where we choose
\be
c_{4A} = c_{4B} = c_{4C} = c_{4D} = c_{4H}=c_{4I} = 0 \, ,
\ee
with only $c_{4E}, \, c_{4F}, \,  c_{4G}, \, c_{4J}, \, c_{4K}$, given by \eqref{ResJ}, non zero.

\section{Conclusion}

Although the results of this paper are rather technical it is important to
understand the consequences of the freedom of the choice of scheme. In
the case of $\cN=1$ supersymmetry the NSVZ expression for the gauge $\beta$-function,
expressed in terms of the anomalous dimension $\gamma$, 
presupposes a particular choice of scheme \cite{JackNSVZ,JackN2}. Results stemming
from the $a$-theorem also constrain $\gamma$ in terms of lower order contributions,
\cite{Freedman}, 
\cite{Barnes}, \cite{Jack}, \cite{Jacka}. The equations used to obtain such results
are not manifestly scheme independent but \cite{Barnes} were able to determine,
for chiral matter fields belonging to a gauge representation $R$, 
the scheme independent contributions $(C_R \, g^2)^p$, $ t_a t_a = C_R \, \mathds{1}$, 
to $\gamma$ for these matter fields for any $p$ from the expansion of
$( 1 - ( 1+ 8\, g^2 \, C_R )^{\frac12})/2$. 
The results quoted above for
the scheme independent $c_1, \, c_2, \, c_{3C}, \, c_{4G}$ can be fitted by
the coefficients in the expansion of $( (1+4x)^\frac12 -1)/4$, but there is 
no derivation of such results to all orders as yet\footnote{Extending the
discussion in \cite{Jack} gives for the planar diagrams 
$2(2 c_{4A} - 2 c_{4C} + c_{4E}) - c_{4G} = \frac32$. The results in \eqref{ResJ}
for $c_{4H}, \, c_{4I}, \, c_{4J}$ may also be obtained as consistency conditions
in terms of $c_{3D}$.}. 
At any order diagrams with a new topology, such as those corresponding to 
$c_{3D}, \, c_{4K}$, so that the contribution to $\tr(\gamma)$ is given by a 
symmetric graph, give rise to new transcendental numbers and so are clearly
independent. 
The presence of $\zeta(3)$ in the scheme invariant in  \eqref{ResJ} containing
$c_{4F}$ involving four loop planar graphs suggests, assuming the results in 
\cite{Avdeev}, \cite{Jack4} are correct, that there is no simple recursive
formula for the coefficients of all graphs with the same basic topology.

One issue which merits further discussion is the role of anomalies since the
irrelevance of antisymmetric contributions  to the anomalous dimension depends
on all global symmetries of the kinetic term being non anomalous. Of course
with chiral fermions this need not be true.

\medskip

\noindent{\bf Acknowledgements}

\vskip 2pt

HO is very grateful to Andy Stergiou for providing the template for
all diagrams in this paper.

\bibliographystyle{utphys}
\bibliography{Scheme}

\providecommand{\href}[2]{#2}\begingroup\raggedright\begin{thebibliography}{10}

\bibitem{Gracey}
J.~A. Gracey, I.~Jack, and C.~Poole, ``{The a-function in six dimensions},''
  \href{http://dx.doi.org/10.1007/JHEP01(2016)174}{{\em JHEP} {\bfseries 01}
  (2016) 174},
\href{http://arxiv.org/abs/1507.02174}{{\ttfamily arXiv:1507.02174 [hep-th]}}.
%%CITATION = ARXIV:1507.02174;%%.

\bibitem{Jack}
I.~Jack and H.~Osborn, ``{Constraints on RG Flow for Four Dimensional Quantum
  Field Theories},''
  \href{http://dx.doi.org/10.1016/j.nuclphysb.2014.03.018}{{\em Nucl. Phys.}
  {\bfseries B883} (2014) 425--500},
\href{http://arxiv.org/abs/1312.0428}{{\ttfamily arXiv:1312.0428 [hep-th]}}.
%%CITATION = ARXIV:1312.0428;%%.

\bibitem{Jack3}
I.~Jack, D.~R.~T. Jones, and C.~G. North, ``{N=1 supersymmetry and the three
  loop anomalous dimension for the chiral superfield},''
  \href{http://dx.doi.org/10.1016/0550-3213(96)00269-6}{{\em Nucl. Phys.}
  {\bfseries B473} (1996) 308--322},
\href{http://arxiv.org/abs/hep-ph/9603386}{{\ttfamily arXiv:hep-ph/9603386
  [hep-ph]}}.
%%CITATION = HEP-PH/9603386;%%.

\bibitem{Jack4}
P.~M. Ferreira, I.~Jack, and D.~R.~T. Jones, ``{The Quasi-infra-red fixed point
  at higher loops},''
  \href{http://dx.doi.org/10.1016/S0370-2693(96)01549-3}{{\em Phys. Lett.}
  {\bfseries B392} (1997) 376--382},
\href{http://arxiv.org/abs/hep-ph/9610296}{{\ttfamily arXiv:hep-ph/9610296
  [hep-ph]}}.
%%CITATION = HEP-PH/9610296;%%.

\bibitem{Avdeev}
L.~V. Avdeev, S.~G. Gorishnii, A.~{\relax Yu}. Kamenshchik, and S.~A. Larin,
  ``{Four Loop Beta Function in the {Wess-Zumino} Model},''
\href{http://dx.doi.org/10.1016/0370-2693(82)90727-4}{{\em Phys. Lett.}
  {\bfseries B117} (1982) 321--323}.
%%CITATION = PHLTA,B117,321;%%.

\bibitem{JackNSVZ}
I.~Jack, D.~R.~T. Jones, and C.~G. North, ``{Scheme dependence and the NSVZ
  Beta function},'' \href{http://dx.doi.org/10.1016/S0550-3213(96)00637-2}{{\em
  Nucl. Phys.} {\bfseries B486} (1997) 479--499},
\href{http://arxiv.org/abs/hep-ph/9609325}{{\ttfamily arXiv:hep-ph/9609325
  [hep-ph]}}.
%%CITATION = HEP-PH/9609325;%%.

\bibitem{JackN2}
I.~Jack, D.~R.~T. Jones, and A.~Pickering, ``{The Connection between DRED and
  NSVZ},'' \href{http://dx.doi.org/10.1016/S0370-2693(98)00769-2}{{\em Phys.
  Lett.} {\bfseries B435} (1998) 61--66},
\href{http://arxiv.org/abs/hep-ph/9805482}{{\ttfamily arXiv:hep-ph/9805482
  [hep-ph]}}.
%%CITATION = HEP-PH/9805482;%%.

\bibitem{Freedman}
D.~Z. Freedman and H.~Osborn, ``{Constructing a c function for SUSY gauge
  theories},'' \href{http://dx.doi.org/10.1016/S0370-2693(98)00649-2}{{\em
  Phys. Lett.} {\bfseries B432} (1998) 353--360},
\href{http://arxiv.org/abs/hep-th/9804101}{{\ttfamily arXiv:hep-th/9804101
  [hep-th]}}.
%%CITATION = HEP-TH/9804101;%%.

\bibitem{Barnes}
E.~Barnes, K.~A. Intriligator, B.~Wecht, and J.~Wright, ``{Evidence for the
  strongest version of the 4d a-theorem, via a-maximization along RG flows},''
  \href{http://dx.doi.org/10.1016/j.nuclphysb.2004.09.016}{{\em Nucl. Phys.}
  {\bfseries B702} (2004) 131--162},
\href{http://arxiv.org/abs/hep-th/0408156}{{\ttfamily arXiv:hep-th/0408156
  [hep-th]}}.
%%CITATION = HEP-TH/0408156;%%.

\bibitem{Jacka}
I.~Jack and C.~Poole, ``{The a-function for gauge theories},''
  \href{http://dx.doi.org/10.1007/JHEP01(2015)138}{{\em JHEP} {\bfseries 01}
  (2015) 138},
\href{http://arxiv.org/abs/1411.1301}{{\ttfamily arXiv:1411.1301 [hep-th]}}.
%%CITATION = ARXIV:1411.1301;%%.

\end{thebibliography}\endgroup

\end{document}